# Loss and fractionation of noble gas isotopes and moderately volatile elements from planetary embryos and early Venus, Earth and Mars

Helmut Lammer[1], Manuel Scherf[1], Hiroyuki Kurokawa[2], Yuichiro Ueno[3], Christoph Burger[4], Thomas Maindl[4], Colin P. Johnstone[4], Martin Leizinger[5], Markus Benedikt[5,1], Luca Fossati[1], Kristina G. Kislyakova[4,1], Bernard Marty[6], Guillaume Avice[7], Bruce Fegley[8], Petra Odert[5]

[1]Space Research Institute, Austrian Academy of Sciences, Schmiedlstr. 6, 8042 Graz, Austria

[2]Earth-Life Science Institute, Tokyo Institute of Technology, 2-12-1 Ookayama, Meguro-ku, Tokyo 152-8550 Japan

[3]Department of Earth and Planetary Sciences, Tokyo Institute of Technology, 2-12-1 Ookayama, Meguro-ku, Tokyo 152-8551 Japan

[4]University of Vienna, Department of Astrophysics, Türkenschanzstrasse 17, 1180 Vienna, Austria

[5]Institute of Physcis/IGAM, University of Graz, Universitätsplatz 5/II, 8010 Graz, Austria

[6]Centre de Recherches Pétrographiques et Geéochemiques, UMR CNRS & Université de Lorraine, 15 Rue Notre Dame des Pauvres, 54501 Vandoeuvre-les-Nacy Cedex, France

[7]Université de Paris, Institut de physique du globe de Paris, CNRS, F-75005 Paris, France

[8]Department of Earth and Planetary Sciences, 1 Brookings Drive, St. Louis, MO 63130, USA

**Abstract** Here we discuss the current state of knowledge on how atmospheric escape processes can fractionate noble gas isotopes and moderately volatile rock-forming elements that populate primordial atmospheres, magma ocean related environments, and catastrophically outgassed steam atmospheres. Variations of isotopes and volatile elements in different planetary reservoirs keep information about atmospheric escape, composition and even the source of accreting material. We summarize our knowledge on atmospheric isotope ratios and discuss the latest evidence that proto-Venus and Earth captured small $H_2$-dominated primordial atmospheres that were lost by EUV-driven hydrodynamic escape after the disk dispersed. All relevant thermal and non-thermal atmospheric escape processes that can fractionate various isotopes and volatile elements are discussed. Erosion of early atmospheres, crust and mantle by large planetary impactors are also addressed. Further, we discuss how moderately volatile elements such as the radioactive heat producing element $^{40}$K and other rock-forming elements such as Mg can also be outgassed and lost from magma oceans that originate on large planetary embryos and accreting planets. Outgassed elements escape from planetary embryos with masses that are $\leq M_{\text{Moon}}$ directly, or due to hydrodynamic drag of escaping H atoms originating from primordial- or steam atmospheres at more massive embryos. We discuss how these processes affect the final elemental composition and ratios such as K/U, Fe/Mg of early planets and their building blocks. Finally, we review modeling efforts that constrain the early evolution of Venus, Earth and Mars by reproducing their measured present day atmospheric $^{36}$Ar/$^{38}$Ar, $^{20}$Ne/$^{22}$Ne noble gas isotope ratios and the role of isotopes on the loss of water and its connection to the redox state on early Mars.

# 1 Introduction

The earliest atmospheric records of Venus, Earth and Mars are stored in the noble gases and their isotopes. The reason is that these elements do not react with other atmospheric species or the surface. Noble gas elements such as Ne, Ar, Kr and Xe and their isotopes contain historical records related to the early evolutionary processes starting during the accretion phase of planetesimals to planetary embryos before and after disk dispersal (Ozima and Podosek, 2002; Porcelli et al., 2002).

Depending on how fast planetary embryos grow to larger masses within the disk and at which orbital locations they originate or migrate, embryos will evolve to objects with different volatile and elemental abundances, which results in different elemental ratios compared to their initial ones (Brasser et al. 2019; this issue). Heating by short-lived $^{26}$Al and $^{60}$Fe radioisotopes shape the thermal history and interior structure of growing planetesimals and planetary embryos during the early stages of planetary formation (e.g., Lichtenberg et al., 2016; Young et al., 2019). The subsequent thermo-mechanical evolution, such as magma ocean-related internal differentiation, rapid volatile degassing, and the possible accumulation of $H_2$/He-envelopes has important implications for the final structure, composition and evolution of terrestrial protoplanets (Ikoma and Genda, 2006; Elkins-Tanton, 2012; Massol et al., 2016; Stökl et al., 2016; Lichtenberg et al., 2016; Benedikt et al., 2019). Under these extreme conditions, noble gases with low condensation temperatures such as Ar and Ne, significant amounts of major rock-forming elements (e.g., K, Si, Mg, Fe, Ca, Al, Na, S, P, Cl), and other moderately volatile elements like Rb or Zn with condensation temperatures of ≈ 1000 K and lower (Lodders et al., 2009) populate the hot accretionary atmospheres to significant amounts (Albarède and Blichert-Toft, 2007; Schaefer and Fegley, 2007; Fegley et al., 2016).

If one considers for instance a mixture of ordinary chondrites with enstatite-like material and/or carbonaceous chondrites with ureilites as recently suggested by Schiller et al. (2018) as indicated by specific isotopic abundances to be the parent bodies of the Earth, then the terrestrial K/U ratio would be higher than the present-day ratio. This indicates that most likely some loss process during proto-Earth's accretion was necessary to deplete moderately volatile elements and fractionate K from U. Since radioactive decay of the heat producing element $^{40}$K contributes to drive a long-lived magnetic dynamo and influences the long-term temperature evolution of a planet's interior (e.g. Turcotte and Schubert, 2002; Murthy et al., 2003; Nimmo, 2015; O'Neill et al., 2019; this issue), it is very important to investigate how atmospheric escape can modify the amount of $^{40}$K and other moderately volatile elements such as Si, Mg, Fe, Ca, Al, Na, S, P, Cl, Rb, Zn until an Earth-like planet finishes its accretion.

Although planetary embryos grow within the protoplanetary disk (Wade and Wood, 2005), these bodies are differentiated and depleted in moderate volatile elements (Hin et al., 2017; Young et al., 2019) due to the loss of a magma ocean-related catastrophically outgassed steam atmospheres (Odert et al., 2018b; Benedikt et al., 2019) as soon as they are not surrounded by the nebula gas anymore. Hydrodynamic atmospheric escape, which is controlled by planetary parameters (e.g. Lammer et al., 2016; Owen and Wu, 2016; Guo, 2019) and the EUV flux of their young host stars (e.g. Koskinen et al., 2013; Erkaev et al., 2013, 2015, 2016; Lammer et al., 2014; Odert et al., 2018b; Kubyshkina et al., 2018c; Guo, 2019; Benedikt et al., 2019; Zahnle et al., 2019), can be very efficient if the mass of the protoplanetary body is small and the main upper atmospheric species is hydrogen. If hydrogen, which either originates from gas accretion in the disk or from dissociation of $H_2O$ vapour of a magma ocean-related outgassed steam atmosphere, escapes efficiently, H atoms can drag along heavier species to space (Zahnle and Kasting, 1986; Hunten et al., 1987; Zahnle et al., 1990; Chassefière, 1996a, 1996b; Odert et al., 2018b; Guo, 2019; Benedikt et al., 2019; Zahnle et al., 2019). The conditions under which this so-called post-nebula atmospheric escape and volatilization processes take place, further lead to the modification of volatile composition and elemental abundances, as well as to the odd elemental and isotopic ratios that are observed in the Solar System planets (e.g. Gillmann et al., 2009; Marty, 2012; Jellinek and Jackson, 2015; Bonsor et al., 2015; Carter et al., 2015; Benedikt et al., 2019; Lammer et al., 2019).

The main aim of this work is to review the latest knowledge on various atmospheric evolution and escape processes that fractionate isotopes and elements during Venus', Earth's and Mars' atmosphere evolution since their accretion phase in the disk. Sect. 2 summarizes our knowledge on the main atmospheric isotopic species and the composition of terrestrial planets obtained from various space missions. In Sect. 3 we discuss the evidence that accreting proto-Venus and Earth captured primordial atmospheres during their growth within the disk. Here we also discuss briefly the new knowledge obtained from exoplanet research and the discovery of $H_2$-dominated low mass exoplanets. In Sect. 4 we briefly review various thermal and non-thermal atmospheric loss processes and discuss their role in isotopic fractionation of different atmospheric species. In Sect. 5 the role of impact erosion of primordial atmospheres and crustal material as well as the role of this process in the modification of volatile elements of accreting protoplanets are discussed. In Sect. 6 we address the loss and isotopic and elemental fractionation of rock-forming volatile elements and noble gases from solidifying magma oceans from Moon- to Mars-like planetary embryos. In Sections 7 and 8 we address the latest attempts for constraining the early evolution of Venus and Earth and Mars by reproducing their present noble gas and elemental ratios until these planets finished accretion. Sect. 9 concludes the review.

## 2 Atmospheric isotopic compositions of the terrestrial planets

Noble gases can be seen as a key to a planet's past because their abundances and isotope ratios represent the earliest record of atmospheres of terrestrial planets. Their evolved isotope ratios contain records of cataclysmic events that the evolving atmospheres experienced. These could be collisions with planetary embryos, comets, mass-fractionating atmospheric escape processes, outgassing from the planetary interior and geologic upheavals (Becker et al., 2003; Baines et al., 2013; Lammer et al., 2018).

Table 1 summarizes and compares the observed atmospheric isotope ratios in the atmospheres of Venus, Earth and Mars. The large Martian atmospheric $^{14}N/^{15}N$ disequilibrium (Fox and Hać, 1997; Füri and Marty, 2015) between the atmosphere and mantle and the similar but smaller disequilibrium on Earth (Cartigny and Marty, 2013) combined with the fractionation of the significantly heavier Xe isotopes (e.g., Pepin, 1991; 2000; Cassata, 2017; Avice et al., 2018; Zahnle et al., 2019; Avice and Marty, 2019; this issue) are a strong indication for different mass- and/or ionization-related atmospheric escape processes and atmosphere-magnetosphere environments of early Earth and Mars. These processes were shaped by the young Sun's radiation and plasma conditions (Lammer et al., 2018; and references therein). A detailed discussion on the nitrogen and Xe disequilibria in Earth's and Mars' atmosphere can be found in Avice and Marty (2019; this issue). Additionally to Earth, icy moons such as Triton or the dwarf planet Pluto also possessing $N_2$-dominated atmospheres. For example, Titan has a 1.45 bar atmosphere with a $N_2$-dominated atmosphere (Strobel and Shemansky, 1982; Coustenis and Taylor, 2008). However, the origin of such atmospheres is related to early photolysis of accreted and outgassed $NH_3$ from subsurface $H_2O$-$NH_3$ oceans, leading to a very different evolution compared to Earth's $N_2$ atmosphere (Coustenis and Taylor, 2008; Mandt et al., 2009, 2014). The origin and evolution of atmospheres that are constrained by $^{14}N/^{15}N$ isotope ratios at frozen worlds like Titan, Triton, and Pluto are discussed in Scherf et al., 2019; this issue). Here we focus mainly on noble gases and moderate volatile rock-forming element evolution on planetary embryos, early Venus, Earth and Mars that originated in an inner planetary system.

The present atmospheric $^{36}Ar/^{38}Ar$ and $^{20}Ne/^{22}Ne$ noble gas isotope ratios can provide historical records on the growth rate of proto-Venus and Earth in the presence of the nebula (Lammer et al., 2019) and hence their early atmospheric evolution, which is also the case for Mars (e.g., Jakosky et al., 1994; 2017; Kurokawa et al., 2018). The atmospheric $^{36}Ar/^{22}Ne$ ratio also contains evidence related to the post-primordial atmosphere impactor composition that was delivered to terrestrial planets (Marty and Allé, 1994; Marty 2012; Lammer et al., 2019).

A reproduction of the primordial and steam atmosphere related D/H ratios yields information that may lead to a better understanding of the oxidation states and initial $H_2O$ inventories of terrestrial planets. The terrestrial water (oceans) and meteorites are enriched by a factor of about 6 in D relative to the protosolar nebula value (Robert et al., 1999; Marty, 2012; Pahlevan et al., 2019). Besides heterogeneities in the disk, mass fractionation during atmospheric escape processes from early terrestrial planets can enrich water in oceans in D, since the lighter isotopes are preferentially concentrated in the escaping upper atmosphere relative to ocean water (Genda and Ikoma, 2008).

The preservation of a carbonaceous chondritic D/H signature with a mean statistical D/H ratio of $(150 \pm 10) \times 10^{-4}$ (e.g., Robert et al., 2012) in Earth's seawater (Marty, 2012) provides evidence that early Earth's ocean underwent minimal D-enrichment via equilibration with $H_2$ followed by hydrodynamic escape. According to the recent results of Pahlevan et al. (2019) the close match of ≈ 10 - 20 % between Earth's seawater and carbonaceous chondrites further constrains the prior existence of an atmospheric $H_2$-inventory of any origin on post-giant-impact Earth to < 20 bars. Moreover, the observed oxidation of silicate Earth occurred before or during the crystallization of the final magma ocean (Pahlevan et al., 2019). Unless D escape is very efficient on Venus, the present H escape flux (averaged over a solar cycle) cannot be larger than about $10^7$ cm$^{-2}$ s$^{-1}$, if the present water vapour in Venus' atmosphere is a remnant of water deposited billion years ago (Donahue, 1999). On the other hand, these authors argued that, if the escape flux remains constant and is as large as about $3\times10^7$ cm$^{-2}$ s$^{-1}$, water would be the remnant of H$_2$O outgassed only about 500 million years ago. If the hydrogen escape fluxes were $> 3 \times 10^7$ cm$^{-2}$ s$^{-1}$ during the past 100 Myrs then the present day D/H ratio in Venus' atmosphere could have been created since a massive resurfacing event a few 100 Myrs ago (Grinspoon, 1993).

Table 1. Atmospheric isotope ratios of the terrestrial planets in the Solar System.

| Atmospheric isotope ratios | Venus | Earth | Mars |
|---|---|---|---|
| D/H | $(160 \pm 20) \times 10^{-4}$ (a) <br> 0.064 – 0.08 (above 70 km altitude; b) | $(1.49 \pm 0.03) \times 10^{-4}$ (c) <br> $1.56 \times 10^{-4}$ (sea water; d) | $(7.58\text{-}10.9) \times 10^{-4}$ (e) <br> $< 1.99 \times 10^{-4}$ (interior; f) |
| $^3$He/$^4$He | Not measured; expected: $< 3 \times 10^{-4}$ (g) | $1.37 \times 10^{-6}$ (h) | Not measured |
| $^{12}$C/$^{13}$C | $88^{(+2,-1)}$ (i) | 89 (inorganic; h) | $85.1 \pm 0.3$ (e, j) |
| $^{14}$N/$^{15}$N | $273^{(+70,-46)}$ (k) | 272 (h) | $173 \pm 11$ (l) |
| $^{16}$O/$^{17}$O | Not measured | 2520 (h) | $2577 \pm 12$ (e, j) |
| $^{16}$O/$^{18}$O | $500 \pm 25$ (k) | 489 (h) | $462 \pm 2.5$ (e, j) |
| $^{20}$Ne/$^{22}$Ne | $11.8 \pm 0.7$ (a) | 9.78 (m) | $10.1 \pm 0.7$ (n) |
| $^{21}$Ne/$^{22}$Ne | Not measured; expected: $< 0.067$ (a) | 0.029 (m) | Not measured (h) |
| $^{33}$S/$^{32}$S | Not measured; expected: $\approx 8 \times 10^{-3}$ (h) | $8.01 \times 10^{-3}$ (h) | Not measured (h) |
| $^{34}$S/$^{32}$S | Not measured; expected: $\approx 0.04$ (h) | 0.045 (h) | Not measured (h) |
| $^{36}$Ar/$^{38}$Ar | $5.6 \pm 0.6$ (o) | $5.32 \pm 0.33$ (p) | $4.2 \pm 0.1$ (n) |
| $^{40}$Ar/$^{36}$Ar | $1.03 \pm 0.04$ (k) <br> $1.19 \pm 0.04$ (i) | $298.56 \pm 0.31$ (q) | $(1.714 \pm 0.17) \times 10^3$ (e, j) |
| $^{136}$Xe/$^{132}$Xe | Not measured; expected: $\approx 1$ (h) | $0.3294 \pm 0.0004$ (r) | $0.3451 \pm 0.0023$ (r) <br> ALH84001: $0.3450 \pm 0.002$ (s) |
| $^{129}$Xe/$^{132}$Xe | Not measured; expected: $\approx 3$ (h) | $0.983 \pm 0.001$ (p) | $2.522 \pm 0.006$ (r) <br> ALH84001: $2.151 \pm 0.032$ (s) |

(a) Donahue et al. (1982), (b) Bertaux et al. (2007), (c) Lécuyer et al. (1998), (d) Michael (1988); (e) Webster et al. (2013a; 2013b), (f) Hallis et al. (2012), (g) von Zahn et al. (1983), (h) Baines et al. (2013), (i) Istomin et al. (1980a), (j) Avice et al. (2018), (k) Hoffman et al. (1980), (l) Wong et al. (2013), (m) Marty (2012), (n) Pepin (1991), Atreya et al. (2013), (o) Istomin et al. (1980b), (p) Ozima and Podosek (2002), (q) Lee et al. (2006), (r) Ozima et al. (1983), (s) Cassata (2017).

Hydrogen escape rates based on the loss of suprathermal H atoms and ions on present Venus were studied by Lammer et al. (2006). These authors modelled photochemically produced suprathermal H atom escape rates in the order of $\approx 3.8 \times 10^{25}$ s$^{-1}$ which is in agreement with the estimates of Donahue and Hartle (1992). Ion escape rates were uncertain at that time and only model-based $H^+$ escape rates in the order of $\approx 7.0\times10^{25}$ s$^{-1}$ largely overestimated the escape from Venus over the planets nightside by assuming ion acceleration by an outward electric polarization force related to ionospheric holes (Hartle and Grebowsky, 1993). A recent study that investigated the $H^+$ escape from Venus during 8 years of data from the Venus Express ion mass analyser instrument obtained much lower $H^+$ ion loss rates in the order of $\approx 7.6 \times 10^{24}$ s$^{-1}$ at solar minimum and $\approx 2.1 \times 10^{24}$ s$^{-1}$ during solar maximum (Persson et al., 2018). These authors found that the decrease of the $H^+$ ion escape from solar minimum to maximum is caused mainly by the change in the flow direction in the magnetotail. During solar maximum a significant Venus ward ion flow close to the planet and its magnetotail was observed (Persson et al., 2018). Thus, the present day Venus total hydrogen escape is in the order of $\approx 4.5 \times 10^{25}$ s$^{-1}$ ($\leq 10^7$ cm$^{-2}$ s$^{-1}$) with suprathermal H atom escape as the dominant process.

The findings of lower $H^+$ ion escape rates by Persson et al. (2018) compared to previous estimates indicates that Venus' present D/H ratio is most likely a remnant of the ancient evaporated water inventory and subsequently outgassed water vapour that were modified by various non-thermal atmospheric escape processes during time periods that lasted longer than the last resurfacing event 500±200 Myr ago.

Other isotopes such as $^{16}O/^{17}O$, $^{16}O/^{18}O$ may reveal information related to a common kinship of the planets (e.g., Clayton, 1993; Yurimoto et al., 2006), $^{33}S/^{32}S$ and $^{34}S/^{32}S$ isotopes contain evidence on the past and current planetary volcanic activity and magmatic composition (e.g., Mahaffy et al., 2012), outgassing from the interior is recorded by $^4He$, $^{40}Ar$ and $^{129,131-136}Xe$ (e.g., Allègre et al., 1987; Marty and Allé, 1994; Watson et al., 2007), and $^{12}C/^{13}C$ can be related to a possible biological activity (e.g., Galimov, 1985; 2003). As one can see in Table 1, several isotopes have large error bars or were not measured by spacecraft in the past. Therefore, new future in-situ measurements of these isotopes as discussed in Dandouras et al. (2019; this issue) in planetary atmospheres are necessary for constraining and understanding the state and evolution of terrestrial planets in the Solar System and beyond.

## 3  The disk phase: $H_2$-dominated envelopes on low mass planets

In the Solar System, there are two well-separated classes of planets: rocky planets (Mercury, Venus, Earth, and Mars with average densities between 3.7 and 5.5 g cm$^{-3}$) and gas/ice giants (Jupiter, Saturn, Uranus, and Neptune with average densities between 0.7 and 1.6 g cm$^{-3}$). The rocky planets have a mass of about one Earth-mass ($M_{Earth}$), or less, while gas giants have masses of tens to hundreds of $M_{Earth}$. However, the largest population of exoplanets discovered to date has a mass/radius falling between those of the Earth and Neptune (e.g., Mullally et al., 2015), demonstrating that the variety of planets existing in nature is much larger than previously thought.

### 3.1 Low mass exoplanets with primordial atmospheres

The mass-radius, or equivalently mass-density, diagram is the first diagnostic enabling the basic classification of a planet, e.g. identify whether a planet hosts a $H_2$-dominated atmosphere or not. The rapidly increasing sample of Neptune-like planets and smaller for which we have a measurement of both mass and radius indicates that these planets present a surprisingly large variety of densities (see Fig. 1).

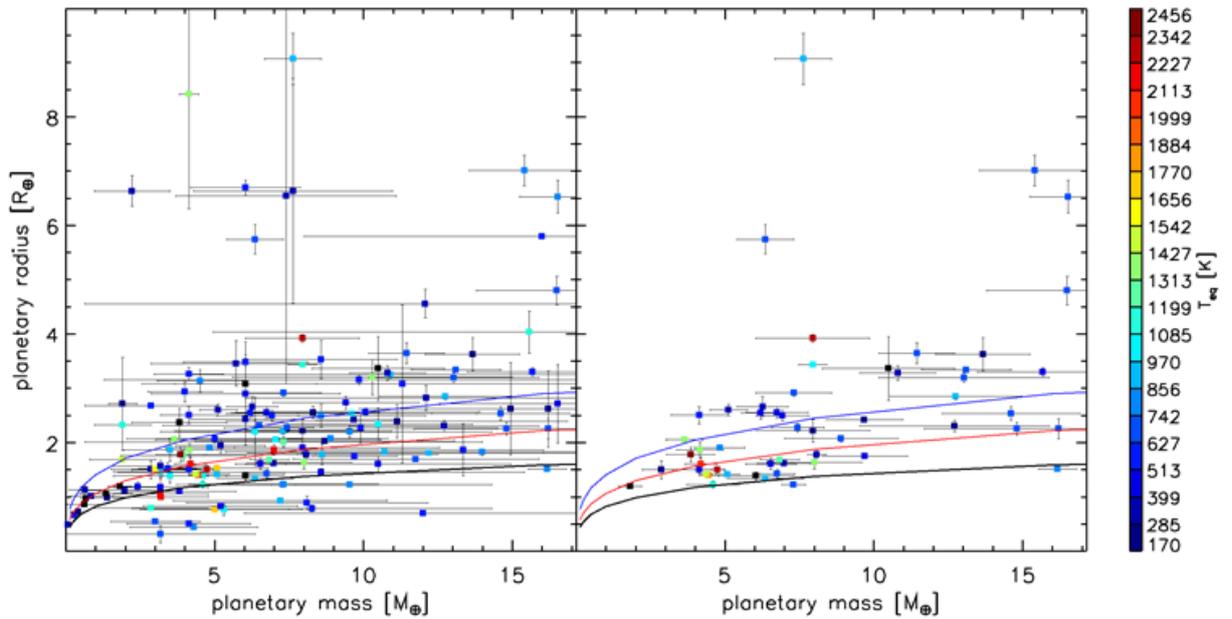

**Fig. 1.** Left: known exoplanets with masses smaller than Neptune in the mass vs radius plane, in Earth units. The color indicates the planetary equilibrium temperature $T_{eq}$. The black, red, and blue lines indicate compositions corresponding to 100% Fe, 100% Rock, and 100% water, respectively (from Zeng et al. 2016). Right: same as left panel, but considering only planets with a mass and radius known to better than 25%. Data from the NASA exoplanet archive (https://exoplanetarchive.ipac.caltech.edu/).

This can be clearly seen in the left panel of Fig. 1 showing the position of the known exoplanets with a mass smaller than Neptune in the mass vs. radius diagram, color-coded by the planetary equilibrium temperature ($T_{eq}$). This plot clearly indicates the presence of a large spread in planetary radius, and thus density, at a given mass, particularly for planets with masses larger than 1.5-2.0 $M_{Earth}$. The same can be seen also when considering only systems for which masses and radii are known to better than 25% (Fig. 1, right panel), indicating that this large spread in bulk density is not the result of large uncertainties on the planetary parameters.

Although planetary cores with various densities/compositions certainly exist, this spread is most likely tightly related to the existence of a large variety of planetary atmospheres that can range from being $H_2$-dominated primordial atmospheres to $CO_2$/$H_2O$-dominated steam atmospheres, or secondary atmospheres, thus more compact. Furthermore, this spread does not seem to be directly correlated with the amount of irradiation they receive from the host star. The origin of this phenomenon is unclear and several factors are likely to play a role, such as protoplanetary nebula lifetime (e.g., Stökl et al., 2015; 2016), mass accretion time (e.g., Schiller et al., 2018), migration history, collisions (e.g., Bonomo et al., 2019), received amount of high-energy irradiation (e.g., Fossati et al., 2017; Owen and Wu, 2017; Zahnle and Catling, 2017; Jin and Mordasini, 2018; Kubyshkina et al., 2018a; 2018b), possible biases in the measurement of planetary radii due to the presence of high-altitude aerosols (e.g., Lammer et al., 2016; Cubillos et al., 2017a; 2017b).

It has been shown that these planets are likely to have Earth-like density cores (Owen and Wu, 2017) and the observations conducted to date clearly show that a number of them still host an $H_2$-dominated atmosphere, most likely accreted from the protoplanetary nebula, despite their close distance to the host star. It is therefore possible that also the inner planets of the Solar System began their evolution with a rocky core surrounded by an $H_2$-dominated envelope.

### 3.2 Evidence of captured nebula gas during early Earth's and Venus' accretion

Besides disk lifetime (e.g. Montmerle et al., 2006; Bollard et al., 2017; Wang et al., 2017) and the young Sun's X-ray, EUV and FUV fluxes and plasma winds (Tu et al., 2015; see also Güdel et al., 2019; this issue), the accretion time is the relevant factor on whether a protoplanet captured enough mass to be able to accumulate and sustain an $H_2$-dominated envelope after the disk disappeared (e.g., Ikoma and Genda, 2006; Lammer et al., 2016; Stökl et al., 2015; 2016; Lammer et al., 2019). In the case of the early terrestrial planets in the Solar System, scientists of various research areas debate since decades whether it was possible for proto-Venus and Earth to accrete enough mass during the disk lifetime for accumulating nebula gas before the disk disappeared. If a protoplanet reached a mass $> 0.5 M_{Earth}$ before the disk evaporated it could have captured a small $H_2$-dominated envelope that remains after the nebula dissipated (e.g., Ikoma and Genda, 2006; Stökl et al., 2015; 2016; see also Lammer et al., 2019b; this issue).

Captured nebula gas was used in past attempts to reproduce the present atmospheric $^{36}Ar/^{38}Ar$ and $^{20}Ne/^{22}Ne$ noble gas ratios of Venus and Earth by various researchers (e.g., Sekiya et al.. 1980a; 1980b; Sasaki and Nakazawa, 1989; Harper and Jacobsen, 1996; Pepin, 2000, Porcelli and Pepin, 2000; Porcelli et al., 2001; Gillmann, 2009; Odert et al., 2018; Lammer et al., 2019). In the case of the Earth, there is data from $^{20}$Ne/$^{22}$Ne isotope ratios in mantle material indicating that a fraction of noble gases has been trapped in Earth's interior in solar composition from the protoplanetary disk (Mizuno et al. 1980; Porcelli et al., 2001; Yokochi and Marty 2004; Mukhopadhyay, 2012). Some researchers argued that these $^{20}$Ne/$^{22}$Ne isotope ratios between 12.52-12.75 could have also been reproduced through implantation of solar wind onto accreted material (Trieloff, 2000; Podosek, 2003; Feigelson et al., 2002; Raquin and Moreira, 2009; Moreira, 2016; Péron et al., 2016; 2017). Later on, such solar wind related noble gas isotopes could then be outgassed during the magma ocean phase. A recent study, however, measured $^{20}$Ne/$^{22}$Ne values up to $13.03 \pm 0.04$ from deep mantle plumes and determined a ratio for the primordial plume mantle of $13.23 \pm 0.22$ (Williams and Mukhopadhyay 2019). These values are only slightly fractionated but comparable with the nebular (solar) ratio. Even if the debate carries on, this can be seen as a robust evidence for a reservoir of nebular gas that is preserved in the deep mantle.

If the accreting proto-Venus and Earth captured $H_2$-dominated envelopes from the disk then these primordial atmospheres were lost during some time after disk dispersal. Below we discuss various atmospheric escape processes, there efficiencies and capabilities for the fractionation of various elements (i.e., isotopes and volatile elements, including rock-forming ones).

# 4 Atmospheric escape processes and their relevance for elemental fractionation

Losses of atmospheric gases to space are caused by a large number of physical processes taking place in the thermosphere, exosphere, and in the case of the Earth in the magnetosphere. These processes are mostly driven by inputs from the Sun in the forms of its high energy radiative spectrum and the solar wind. For this reason, loss processes were more important during earlier phases of the solar system when the Sun was more active.

## 4.1 Upper atmospheres

The upper atmospheres of planets are of primarily importance for determining how rapidly atmospheric gases are lost to space and for determining the relative losses of different elements and isotopes. Understanding how Earth's upper atmosphere has evolved is important for understanding the effects of losses on isotopic fractionation. Various atmospheric layers of interest are outlined in Fig. 2. The main layers in a planetary atmosphere can be defined by the temperature structure. On Earth, above the troposphere and stratosphere, starting at approximately 50 km altitude, is a region called the mesosphere; in this region, the temperature decreases with altitude due to cooling by $CO_2$ molecules emitting infrared radiation to space.

Above this is the thermosphere, starting at approximately 100 km altitude, which is heated by the Sun's X-ray, extreme ultraviolet and ultraviolet radiation, and therefore has a positive temperature gradient. The top of the thermosphere is the exobase (typically at altitudes of ≈ 500 km for the modern Earth), which is the point above which the gas density is so low that the mean free paths of particles are long enough compared to the scale height of the atmosphere that the gas is non-collisional. This non-collisional region above the exobase is the exosphere. An additional important layer is the thermosphere-ionosphere region, which starts in the upper mesosphere and extends to the exobase. This is the domain in which most of the Sun's X-ray and extreme ultraviolet radiation is absorbed, which causes photoionization of the atmospheric constituents.

In the thermosphere, the composition of the gas not only changes due to molecular diffusion, but also due to photodissociation and photoionization, driven by the absorption of solar radiation at wavelengths below approximately 200 nm. For the modern Earth, the dissociation of $O_2$ and $N_2$ molecules, combined with molecular diffusion, mean that the gas in the exobase is primarily composed of atomic O and N. The absorption of this radiation has several additional effects on the gas, with the most important being heating. When the atmosphere is in a steady state, this heating of the upper atmosphere is balanced by cooling from infrared emission to space, particularly by $CO_2$.

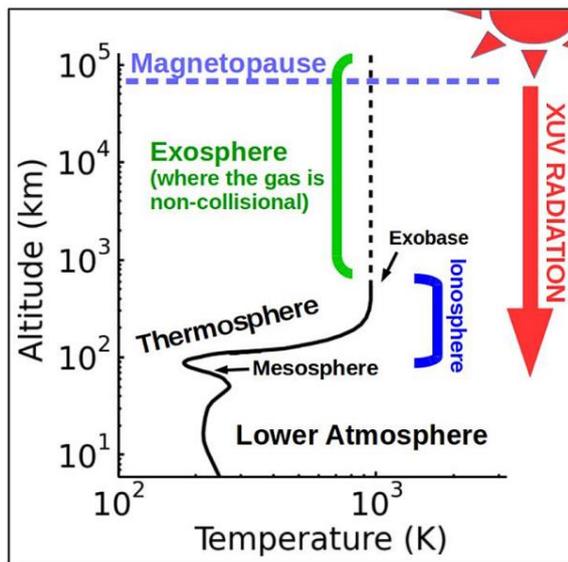

**Fig. 2**. Earth's atmospheric temperature structure and regions of interest.

Locally within the gas however, heating and cooling do not balance, and downward thermal conduction is in fact the main process responsible for cooling the upper thermosphere.

For Venus and Mars, the same basic set of physical processes are important in the upper atmospheres, with the only additional important process being heating by the absorption of solar infrared radiation by $CO_2$ (Fox et al., 1991). The high $CO_2$ concentration in these atmospheres also means that infrared cooling is much stronger and the thermospheres are therefore much cooler. The higher concentrations of $CO_2$ expected in the Earth's atmosphere at earlier times would also likely have this effect, cooling the atmosphere and reducing the effects of atmospheric loss (Kulikov et al., 2007; Johnstone et al., 2018).

It is common to break down the vertical structure of an atmosphere into the homosphere and heterosphere. In the homosphere, which for the Earth extends from the surface to the homopause at approximately 120 km altitude, turbulent mixing of the gas causes the mixing ratios of all long lived species (those that are not rapidly created and destroyed by chemical processes) to be uniform with altitude regardless of the particle masses.

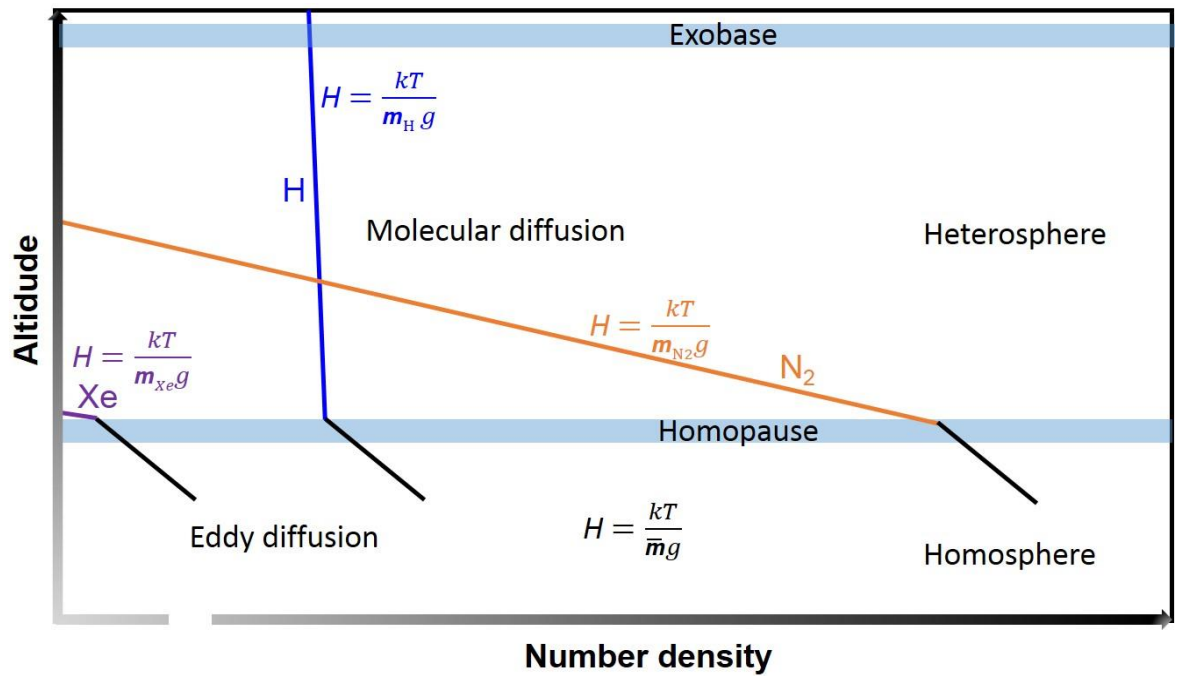

**Fig. 3.** Illustration of density-altitude structures in the homosphere-heterosphere transition region. In the homosphere the atmosphere is mixed and all species are distributed by the same average scale height (homosphere), while above the homopause all species are distributed by their own scale heights as a function of mass. The density of heavy elements decreases very fast compared to light elements.

This process is often known as eddy diffusion and is responsible for allowing heavier particles such as $CO_2$ molecules and noble gases such as Ne, Ar, and even Xe atoms to reach the lower thermosphere. Without eddy diffusion, the abundances of heavier particles would rapidly drop with increasing altitude as molecular diffusion separates the particles by mass. This is indeed what happens above the homopause, with the abundances of heavier particles becoming extremely low in the mid and upper thermospheres. This process contributes to causing lighter particles to be lost from the atmosphere more efficiently than heavier particles.

A related and imported concept is that of diffusion-limited escape. This is itself not a loss process but an upper limit on how rapidly minor (low-abundance) species can be lost from the upper atmosphere of a planet. This limit exists because minor species cannot be lost from the upper atmosphere more rapidly than they can be transported upwards by molecular diffusion at the homopause. The reason to consider the homopause here, or at least the region directly above the homopause, is that this is the region in the heterosphere where molecular diffusion is the least efficient due to the high density and relatively low temperature.

For instance, H atom escape on present Earth, Venus and Mars is limited by upward diffusion through the upper atmosphere (e.g. Hunten, 1975; Kasting and Pollack, 1983). On Earth, $H_2O$, $H_2$, and $CH_4$ move upward in the atmosphere via turbulent mixing up to the homopause. At ≈17 km, the cool temperatures at the tropopause freeze out most of the water vapour, preventing the upward mixing of some hydrogen. Near the homopause hydrogen bearing molecules (i.e. $CH_4$) are dissociated by UV photons so that atomic and molecular hydrogen particles are produced. These H and $H_2$ particles diffuse upward through the upper atmosphere to the exobase level from where they can escape by thermal Jeans escape. On present Earth, the "bottleneck" for hydrogen escape is therefore diffusion-limited. This diffusion-limited flux is proportional to the hydrogen-mixing ratio at the homopause, which is ≈5.0 ×$10^{-7}$ at present Venus (Kasting and Pollack, 1983) and ≈3.0 ×$10^{-5}$ at present Mars (Krasnopolsky and Feldman, 2001). For such low water vapour mixing ratios at the homopause the escape of hydrogen at the exobase level at Mars and Venus, but also at Earth are thermal Jeans escape.

Higher homopause water mixing ratios lead to higher escape rates that are not controlled by diffusion through the upper atmosphere anymore. Kasting and Pollack (1983) found that for $H_2O$ mixing ratios at the homopause that are ≥ 5.0 ×$10^{-4}$ the stellar/solar EUV flux that is absorbed above the homopause level in a hydrostatic atmosphere becomes larger than the energy that is carried away by the thermally escaping particles. At such high $H_2O$ mixing ratios at the homopause level the hydrogen-dominated upper atmosphere becomes unstable against expansion and Jeans escape transfers to hydrodynamic escape. According to Kasting and Pollack (1983), for $H_2O$ mixing ratios at the homopause that are ≥ $10^{-3}$ the hydrogen escape flux reaches values that are comparable to energy limited escape. High $H_2O$-mixing ratios at the homopause level in the order of 0.1 and higher can originate in steam atmospheres (e.g., Lebrun et al., 2013; Massol et al., 2016; Salvator et al., 2017; Benedikt et al., 2019). Under such conditions thermal escape of H atoms from the upper atmosphere and the corresponding fractionation of isotopes and other heavier minor and/or trace species is then controlled by thermal hydrodynamic escape (e.g., Hunten, 1973; Zahnle and Kasting, 1986; Hunten et al., 1987; Zahnle et al., 1990; Odert et al., 2018; Lammer et al., 2019).

**Atmospheric Loss Processes and their influence on isotope fractionation**

Loss processes are often broken down into thermal and non-thermal processes. Thermal processes are related to the thermal energy of the gas and can be primarily broken down into Jeans escape and hydrodynamic escape. Non-thermal processes are essentially all others. Most escape processes preferentially remove lighter isotopes, though they differ in how efficiently they fractionate by mass. The ability of an escape process to fractionate isotopes is often described by a fractionation factor *f*, with *f>1* meaning that the heavier isotope is preferentially removed, *f<1* meaning that the escape process preferentially removes the lighter isotope, and *f=1* describing the case of no fractionation. The Rayleigh distillation relationship is commonly used to describe fractionation of an isotope ratio due to a separate process. This relationship connects the initial, $n_0$, and current, *n*, inventory of the constituent and a fractionation factor of the escape process, *f* (e.g., Lunine et al., 1999; Donahue, 1999; Mandt et al., 2009; 2014; 2015):

$$n_0/n = (R/R_0)^{1/1-f}, \qquad (1)$$

where *R* is the current ratio of the heavy to light isotope abundance, and $R_0$ is the initial ratio. The ratio $R/R_0$ defines the degree of enrichment in the heavier isotope in comparison to the primordial value. The upper limit of the enrichment of the heavier isotope can be found as

$$R/R_0 \leq (\varphi t/n+1)^{(1-f)}, \qquad (2)$$

where *φ* is the maximum escape flux and *t* is time. Different escape processes have various possible maximum escape fluxes and different fractionation coefficients that can depend on the structure and the composition of the upper atmosphere. Below, we briefly summarize the main atmospheric escape processes and their influence on the atmospheric fractionation.

*Jeans escape:* when an upper atmosphere is hydrostatic and is not able to flow away from the planet hydrodynamically, it is still possible for particles in the high-speed tail of the thermal distribution of particles at the exobase to have speeds exceeding the escape velocity. The loss of such particles is known as Jeans escape. This process leads to highly mass-dependent fractionation. For the present solar system planets, Jeans escape rates are quite low, so that it takes a long time to fractionate a large atmospheric reservoir assuming present escape rates (Mandt et al., 2014) though the process would have been much more rapid at earlier times when the Sun was much more active. The fractionation factor *f* for Jeans escape can be written as (e.g., Mandt et al., 2009)

$$f = \sqrt{\frac{m_1}{m_2}} \left[ e^{(\lambda_1 - \lambda_2)} \frac{(1+\lambda_2)}{(1+\lambda_1)} \right]. \qquad (3)$$

*Hydrodynamic escape:* if the upper atmosphere is heated by the EUV flux of the planet's host star to very high temperatures, it is possible that the gas in the upper atmosphere flows away from the planet with a speed at the exobase exceeding the escape speed. Such a flow has the form of a transonic Parker wind (e.g. Tian et al., 2005; Johnstone et al., 2019). Whether an atmosphere reaches this state depends on the gas temperature in the upper atmosphere and the mass of the planet, with hydrodynamic escape taking place significantly more readily for lower mass planets. The source of the energy that heats the gas is typically the X-ray and ultraviolet field of the star. Very early in a planet's history, immediately after the dissipation of the primordial nebula, it is also possible for hydrodynamic escape to be powered by the initial accretion energy of the envelope gas (Stökl et al., 2015) and the bolometric radiative emission of the star (Owen and Wu, 2016), though the latter is unlikely to have been significant for the young solar system planets due to their large distances from the Sun. Previous works have found that hydrodynamic escape removes all isotopes very efficiently, which leads to small fractionations (Mandt et al., 2009). For hydrodynamic escape the fractionation factor f converges towards

$$f = \sqrt{\frac{m_1}{m_2}}, \quad (4)$$

where $m_1$ and $m_2$ are the masses of the lighter and heavier isotope, respectively.

When hydrogen atoms escapes hydrodynamically, however, it will drag heavier elements with it. This effect can subsequently lead to a fractionation of lighter isotopes to heavier isotopes (e.g., Zahnle and Kasting, 1986; Hunten et al., 1987; Zahnle et al., 1990; Odert et al., 2018). If besides hydrogen one assumes a second major species, e.g. oxygen, in the hydrodynamically escaping atmosphere, which is also dragged by hydrogen, then the fractionation factors $x_i$ for the second major species and $x_j$ for an arbitrary amount of additional minor species can be written as (e.g., Odert et al., 2018)

$$x_i = 1 - \frac{g(m_i - m_H)b_{H,i}}{F_H k_B T (1 + f_i)}, \quad (5)$$

and

$$x_j = \frac{1 - \frac{g(m_j - m_H)b_{i,j}}{F_H k_B T} + \frac{b_{H,j}}{b_{H,i}} f_i (1 + f_i) + \frac{b_{H,j}}{b_{i,j}} f_i x_i}{1 + \frac{b_{H,j}}{b_{i,j}} f_i}, \quad (6)$$

where $g$ is the gravitational acceleration of the planet, $k_B$ the Boltzmann constant, $T$ the upper atmosphere temperature, $m_i$, $m_H$, and $m_j$ the masses of the masses of the respective atoms, $b_{H,i}$, $b_{O,j}$, $b_{H,j}$ the binary diffusion parameters, $F_H$ the hydrodynamic escape flux of hydrogen atoms, and $f_i$ the mixing ratio of the second major species to hydrogen.

*Ion pick-up:* above the exobase, neutral atmospheric particles follow ballistic trajectories. If these particles reach the region above the magnetopause or ionopause, they can directly interact with the solar wind, where they can be ionized by solar XUV radiation and interactions with the solar wind. These ions are picked up by the solar wind and are lost from the planet. If a planet does not have an intrinsic magnetic field, as is the case for modern Mars and Venus, then the solar wind can directly interact with the ionosphere of the planet, thus making the ion pick up escape more efficient. For a magnetized planet, atmospheric particles have first to reach the location of the magnetopause to be picked up by the solar wind, which can reduce the ion pick up escape rates in comparison to the non-magnetized case. Due to diffusive separation of atmospheric particles by mass above the homopause (see also Fig. 2), isotopes with lighter masses reach higher altitudes and will therefore be preferentially removed from the top of the atmosphere. For non-thermal escape such as ion pick-up, the fractionation factor can be defined as follows (Lunine et al. 1999, Mandt et al. 2014), i.e.

$$f = exp\left[-\frac{g(\Delta z)(m_2 - m_1)}{k_B T}\right], \tag{7}$$

where $\Delta z = r_e - r_h$ is the distance between the homopause $r_h$ and the exobase $r_e$ levels (see also Fig. 2) if the planet has no intrinsic magnetic field, otherwise the exobase can be substituted by the magnetopause.

*Sputtering:* in this process, atmospheric atoms or molecules are ejected due to collisions between solar wind protons and backscattered pick-up ions with the planetary atmosphere (e.g., Johnson, 1990). This process efficiently fractionates the isotopes through diffusion as a function of the temperature in the upper atmosphere and the distance between the exobase and the homopause. Like with other loss processes, the efficiency of the fractionation by sputtering depends on the exact structure of the upper atmosphere, with expanded atmospheres fractionated more easily (Johnstone et al., 2018). Sputtering preferentially removes the lighter atom or particle. On low mass planetary bodies such as Mars, ion pick up and sputtering are likely the two fractionation processes that led to the observed $^{38}Ar/^{36}Ar$ fractionation (Jakosky et al., 2017). The fractionation factor *f* can be calculated in the same as for ion pick-up escape.

*Plasma instabilities:* plasma instabilities, such as Kelvin-Helmholz instability, can drive an escape of plasma vortexes from the ionospheric boundary from planets which have no intrinsic magnetic dynamo (i.e., Venus, Mars,…). A wave-like structure of initially small amplitudes can grow and form vortexes. These vortexes detach from the ionized upper atmosphere and escape from the planet in a plasma cloud. This escape process is relevant for planets with weak or no intrinsic magnetic fields (Terada et al., 2002; Amerstorfer et al., 2007). Since plasma instabilities remove a large number of ions at once, they are not very efficient isotope fractionators.

*Cold ion-outflows:* ions and free electrons produced by the absorption of solar extreme ultraviolet and X-ray photons are able to escape by a separate process. The expansion of the lighter electrons produces an ambipolar electric field that accelerates the ions away from the planet causing an outflow of ions and electrons (Ganguli, 1996). For magnetised planets like the Earth, this flow takes place at the magnetic poles. This process can accelerate the planetary ions and, if they reach a sufficiently high velocity, lead to them escaping the planet's gravity (e.g., Hartle et al., 1993; Lundin et al., 2011). Heavier ions can also be dragged by other atmospheric constituents through this process, as was probably the case for the easy ionized $Xe^+$ isotopes in the Earth's and Mars' atmosphere (e.g., Zahnle et al., 2019; Avice and Marty, 2019; this issue).

*Photochemical escape:* this type of escape is related to chemical reactions that produce hot atoms with high energies, so called suprathermal or "hot" atoms. If the energy of an atom is high enough, it can exceed the energy of the gravitational binding to the planet, and then the atom is lost (e.g., Amerstorfer et al., 2017). The reactions can include various chemical reactions, including dissociative recombination or charge exchange. Photochemically produced suprathermal atom escape can preferentially remove lighter isotopes. On Mars, photochemical processes lead to removal of nitrogen and oxygen atoms, and similar to the fractionation by ion pick-up, the fractionation is driven by the diffusive separation of different isotopes by mass above the homopause (e.g., Jakosky et al., 1994).

Not all escape processes are equally efficient at causing mass fractionation. For example, hydrodynamic outflow is relatively inefficient at fractionating of elements and isotopes with a small mass difference (i.e., D/H…), while it can be efficient if a light species such as atomic hydrogen drags heavier ones, whereas Jeans escape is very efficient but much less efficient for the total atmospheric escape. As a rule of thumb, high-rate loss processes, such as hydrodynamic escape, are very efficient in removing the atmospheric gas, but are inefficient fractionators in case of light elements where it takes a long time to change the isotopic ratio. On the other hand, low-rate loss processes are efficient fractionators, but take a very long time to influence a large inventory.

*Impact erosion:* early in the planet's history, atmospheric gas can be removed by impacts Due to a large mass of gas removed simultaneously; this process is not an efficient fractionator of isotopes related elements in atmospheres and is discussed in more detail in the subsect. 5.3.

## 5   Impact erosion of early crust, mantle and atmosphere

Understanding impact and collision processes and their consequences during accreting planetary embryos and protoplanets is a vital component in the overall picture, since those events happened during all stages of terrestrial planet formation. The currently established picture distinguishes several main stages, where theoretical knowledge about the early phases from dust to planetary embryos is still limited and various competing theories are discussed (e.g. Raymond et al. 2014; Johansen and Lambrechts, 2017; Izidoro and Raymond, 2018; Lammer et al. 2019a; this issue). However, it is relatively well established that the final phase of accretion, lasting for millions to several 10s of Myr is marked by giant collisions among planetary embryos and remaining planetesimals, which shape the bulk properties of eventually formed planets.

A central aspect of collisional processes in an active planet formation environment is the deposition of large amounts of heat. Together with the release of gravitational energy during differentiation and from short-lived radioactive isotopes, this can lead to largescale melting of outer silicate layers, with the result of local or even global magma oceans. Their general evolution and particularly their solidification time are strongly coupled to the overlying atmosphere, which can consist of primordial nebular gas and/or outgassed volatiles from the current or an earlier magma ocean phase itself.

### 5.1 Impact related atmosphere erosion

As discussed above, growing bodies accrete nebular gas once they become sufficiently massive, where cores larger than one Earth-mass accrete an $H_2$-dominated envelope that most likely remain during their whole lifetime (Johnstone et al., 2015; Owen and Wu, 2016), while protoplanets below $0.5 M_{Earth}$ probably loose most of their primordial atmosphere, which is largely boiled off after the surrounding gas dissipates and potentially replaced by an outgassed one later (Stökl et al., 2016). It is important to mention that existing work on the interplay of internal and external (impact) heating, magma ocean and (proto-)atmosphere evolution of growing terrestrial planets has focused on roughly Earth-mass objects, even though smaller bodies – the size of planetary embryos and below – are the stepping stones on the way to fully grown planets.

Studies on impact effects onto atmosphere and ocean-covered protoplanets were often based on one-dimensional hydrodynamic codes (Genda and Abe, 2003, 2005; Inamdar and Schlichting, 2016), analytical models in combination with such codes (Schlichting et al., 2015), and also 3-dimensional hydrocode simulations of impacts of small bodies (Shuvalov, 2009) up to head-on collisions in the Earth- to super-Earth mass range (Liu et al., 2015). An important result is that smaller impactors are - per unit mass - much more efficient in eroding atmosphere than larger ones (e.g. Schlichting et al., 2015; Schlichting and Mukhopadhyay, 2018, for an overview). The mass loss efficiency (the ratio of lost atmospheric to impactor mass) is largest for small bodies, e.g. in the kilometer range for current Earth, and decreases with mass by orders of magnitude up to the giant impact regime (∼ 1000 km bodies) This is mainly because for small impactors the majority of the energy is available for driving of atmosphere locally; above the tangent plane of the impact location, the gas can be entirely lost.

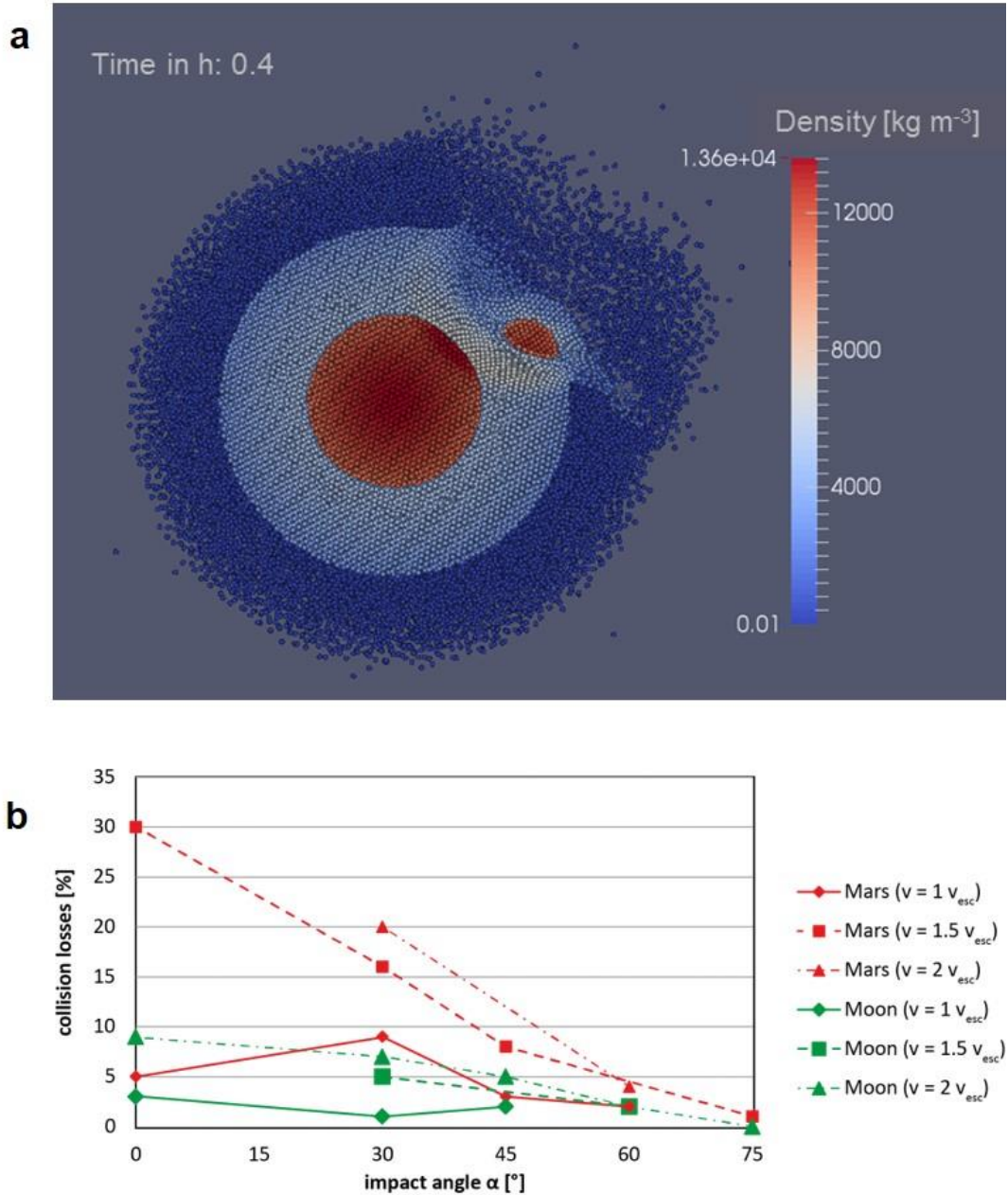

**Fig. 4.** Panel a: Snapshot of a SPH collision simulation between a Moon-mass embryo and a $0.55M_{Earth}$ protoplanet that is surrounded by a primordial $H_2$-envelope, with an impact angle of ≈ 30 degrees and an impact velocity of $2v_{esc}$. The colour coding shows the density and thus the core-mantle-atmosphere structure. Panel b: Atmospheric collision losses in percent of $H_2$-envelope covered protoplanets caused by Moon- and Mars-mass impactors as a function of impact angle and velocity in units of $v_{esc}$.

For impactors with energy deposition larger than what is required for ejecting all material above the tangent plane, considerable parts of their kinetic energy have to be used to drive a strong shock through the target body to eject atmosphere globally, once the shock-induced ground velocity is sufficient to accelerate atmosphere components to escape velocity. The presence of oceans on the surface has been found to aid the escape of overlying atmosphere, due to their lower shock impedance compared to a solid surface (Genda and Abe, 2005). In recent 3-dimensional smoothed-particle hydrodynamics (SPH) giant impact simulations, atmosphere losses of $\approx 5$ % per collision for Moon-mass/size, and $\approx$ 10-20 % for Mars-mass/size impactors in collisions with a $\approx$ 0.5-0.75$M_{Earth}$ protoplanet with primordial atmospheric mass fractions of $\geq$ 0.001 have been obtained (Lammer et al., 2019).

The impact scenarios in this study cover the full range of impact angles and common impact velocities from 1 to 2 × $v_{esc}$. An SPH collision snapshot of a Moon-mass impactor that crashes into an $H_2$-covered protoplanet with 0.5$M_{Earth}$ is shown in panel (a) in Fig. 4 (Burger et al., 2018; Lammer et al., 2019). Panel (b) in Fig. 4 shows eroded primordial $H_2$-dominated atmospheres in percent from protoplanets with masses $\approx$ 0.5 - 0.75$M_{Earth}$ caused by Moon-mars and Mars-mass impactors as a function of impact angle and escape velocity (Lammer et al., 2019). The different results shown in Lammer et al. (2019) indicate that studying atmospheric losses caused by giant impacts only by assuming head-on scenarios (e.g. Liu et al., 2015) may be insufficient to understand the full diversity and efficiency of atmospheric losses by such events. These results confirm that - like for small bodies - even in the giant impact regime lower-mass bodies are more efficient (normalized to impacting mass) than larger ones in eroding atmosphere (Schlichting et al., 2015; Schlichting and Mukhopadhyay, 2018).

### 5.2 Collisional erosion of crust/mantle material by large impactors

During its accretion phase, in particular during the so-called giant impact phase, a growing protoplanet is expected to experience energetic collisions with other planetary embryos (e.g., Quintana et al., 2016). Such a collision can then remove part of the newly formed crust from the planet which will either re-accrete or be lost to space (Carter et al., 2018). This process will therefore preferentially remove incompatible lithophile elements from the growing planet leading to a fractionation between siderophile and lithophile elements (O'Neill and Palme et al., 2008; O'Neill et al., 2019; this issue). While Fe/Mg or Fe/Si, of which Fe is siderophile and Mg and Si are lithophile, will be fractionated by collisional erosion, other ratios such as K/U, which are both lithophile elements, will not. This process could therefore account for the high terrestrial Fe/Mg ratio of 2.1 ± 0.1 compared to the solar ratio of 1.9 ± 0.1 (O'Neill et al., 2008). Another ratio that was suggested to be fractionated by collisional erosion is Earth's $^{142}$Nd/$^{144}$Nd ratio (Jellinek and Jackson, 2015). Here, however, a more recent study found that $^{142}$Nd/$^{144}$Nd might not have been fractionated at all (Bouvier and Boyet, 2016).

Whether collisional erosion can account for the high terrestrial Fe/Mg ratio was numerically studied by Carter et al. (2015) and Bonsor et al. (2015). Even though they both conclude that collisional erosion can fractionate Earth's Fe/Mg from a chondritic to the present-day terrestrial value, they also found that this fractionation can only be reached if Earth was an extreme case. The growing planetary embryos that exhibit a different range of core mass fractions through giant impacts must have had a significant dynamical excitation to finally create the present-day day Fe/Mg ratio of Earth (Bonsor et al., 2015; Carter et al., 2015). A more recent study by Carter et al. (2018) found that a growing protoplanet on average can lose about one third of its crust through collisional erosion which in turn would lead to the loss of up to 20% of its heat producing elements if the crust does not re-accrete onto the planet. Such an alteration of the planetary heat budget might further also have implications on the habitability of a planet since this could strongly affect the evolution of tectonic activity (Jellinek and Jackson, 2015; O'Neill et al., 2019; this issue).

Collisional erosion could further also have been important in view of Mercury's high iron abundance, as was recently simulated by Chau et al. (2018). These authors found that an energetic impactor could alter the iron distribution within a planet by up to 25%. According to their study, Mercury's abnormally high abundance of iron could therefore be the result of multiple collisions. Another recent study by Clement et al. (2019), however, concludes that, even though planets with elevated core fractions were produced in 90% of their simulations through the fragmentation of collided bodies, there is only a probability of ≤ 1% to simultaneously reproduce Mercury and Venus after such an event.

### 5.3 Elemental fractionation caused by large impact events

To summarize the impact erosion processes, one finds two scenarios where large impactors will change the composition of a growing proto-planet. Both cases are illustrated in Fig. 4. If a proto-planet accreted > 0.5 $M_{Earth}$ inside the disk, elements in solar abundance and magma ocean-related outgassed moderately volatile rock-forming elements and noble gases will be embedded as trace elements in a captured $H_2$-dominated primordial atmosphere.

As illustrated in Fig. 5a large impactors that mainly affect the atmosphere can then contribute to the fractionation of moderately volatile elements such as K and/or Mg compared to U or Fe, which remain in the crust/mantle. If a large impactor crashes into the body, so that crustal /mantle material is also eroded (see Fig. 5b), then refractory elements (e.g., U, Th) will also be lost (e.g. Palme and O'Neill, 2003; see also O`Neill et al., 2019; this issue).

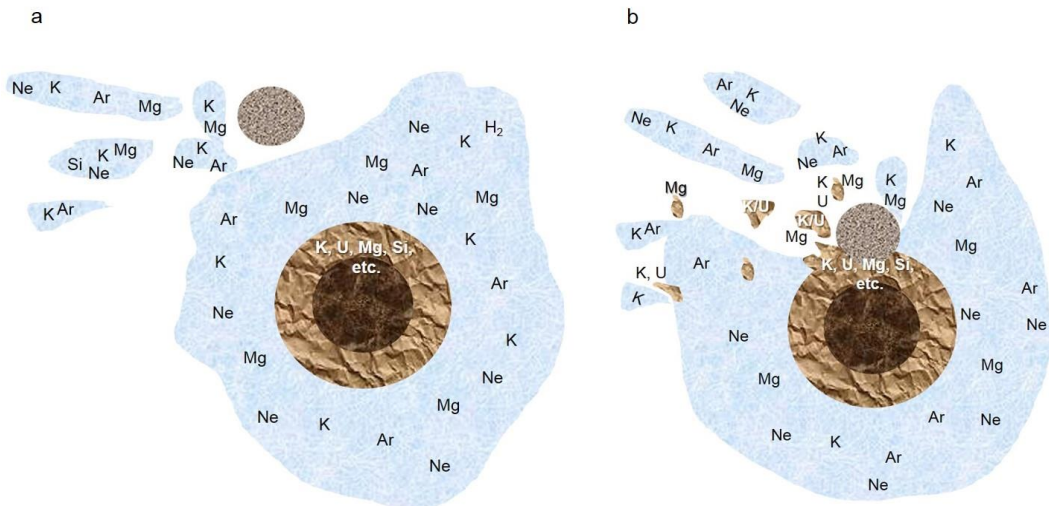

**Fig. 5.** Illustration of the two impact related erosion scenarios. a) Impact erosion of atmospheres, and b) direct impacts and related erosion of crustal and mantle material. Impacts that mainly affect a primordial atmosphere can contribute to the modification of ratios between moderately volatile rock-forming elements that can also be in a primordial atmosphere and refractory elements that remain in the surface, such as U or Th (panel a), while direct impacts can also erode crustal and mantle material (panel b).

Impacts, however, will not fractionate isotopes of atmospheric noble gases (e.g. $^{36}Ar$, $^{38}Ar$, $^{20}Ne$, $^{22}Ne$, etc.) if they only affect the atmosphere (Fig. 5a) but contribute to a depletion of volatile elements from a planetary body. If impactors with different isotopic compositions and ratios are involved in the accretion of a protoplanet (Fig. 5b), noble gas isotope ratios, however, can be modified by their mixing with compositions, abundances and ratios of the impacting bodies as discussed in Sect. 6 Below we discuss atmospheric escape and fractionation of such elements from planetary embryos, which experience magma oceans and therewith related outgassing of volatile elements.

## 6 Escape of rock-forming elements and volatiles from planetary embryos

Terrestrial planets accrete mass by numerous collisions between small objects that accumulate to larger ones, including Moon and Mars-mass planetary embryos that accrete onto full planets (e.g., Johansen and Lambrechts, 2017; Izidoro and Raymond, 2018; Lammer et al., this issue 2019a). Growing planetesimals, which are embedded in the circumstellar disk, with a size between hundred to thousand kilometers develop global magma oceans, caused by the heating of short-lived radioactive elements, like $^{26}Al$ (e.g., Lichtenberg et al., 2016; 2018; Young et al., 2019), and mutual collisions (e.g., Safronov and Zvjagina, 1969; Wetherill, 1980; Tonks and Melosh, 1993). Schaefer and Fegley (2007) and Fegley et al. (2016) performed chemical kinetic and equilibrium calculations for modeling the chemistry of moderate volatile elements (e.g., K, Si, Mg, Fe, Ca, Al, Na, S, P, Cl, F) released by heating of different types of carbonaceous, ordinary and enstatite chondrites as a function of temperature and pressure of early atmospheres. These studies yield significant amounts of major rock-forming and other volatile elements, such as Rb and Zn, and noble gases (Ar, Ne, etc.) in primordial and/or catastrophically outgassed steam atmospheres at high temperatures (1000-2500 K).

When the disk dissipates, after a short but efficient boil-off phase (Owen and Wu, 2016; Lammer et al., 2018) the magma oceans of the former disk-embedded bodies solidify and steam atmospheres are catastrophically outgassed (see Fig. 6). Depending on the redox state these catastrophically outgassed atmospheres contain $H_2$, CO (reduced interior) and/or $H_2O$ and $CO_2$ (oxidized interior) (Nikolaou et al., 2019). The outgassed amount per element depends on the bulk composition of a particular planetary embryo and the solidification process of the mentioned magma ocean (e.g., Elkins-Tanton, 2008; 2012; Salvador et al., 2017; Nikolaou et al., 2019).

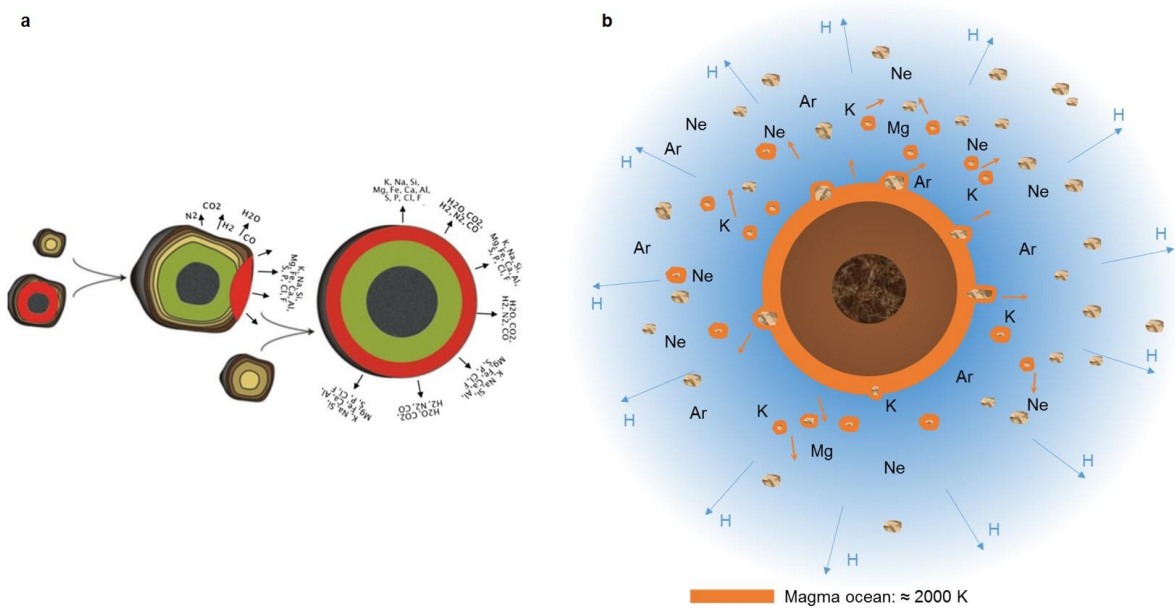

**Fig. 6.** a) Illustration of planetary embryo growth due to collisions between differentiated planetesimals/planetary embryos. The red areas correspond to magma from which various elements evaporate and can be lost directly at bodies with masses ≤ $M_{\text{Moon}}$ (Benedikt et al., 2019; Young et al., 2019) and by dragging of hydrodynamically escaping H atoms dissociated either from $H_2O$ molecules originated from catastrophically outgassed steam atmospheres during magma ocean solidification (Benedikt et al., 2019) or b) from remnants of captured $H_2$-dominated nebula-based envelopes (Lammer et al., 2019).

Recently, Benedikt et al. (2019) studied the losses of magma ocean-related catastrophically outgassed steam atmospheres and outgassed rock-forming elements as well as Ar and Ne noble gases from planetary embryos with masses between $1.0 M_{Moon}$ and $1.5 M_{Mars}$ along the EUV-flux evolution tracks of slow and moderate/fast rotating young G-type stars (Tu et al., 2015; Güdel et al., 2019; this issue) at orbital distances between 0.5 – 1.5 AU.

Based on studies from Wakita and Schmitt (1970), Right and Chabot (2011), and Elkins-Tanton (2012), one can assume magma ocean depths for planetary embryos within the mass ranges mentioned above between 500 – 1000 km for surface temperatures of 1500 – 3000 K. Because of the low mass of planetary embryos and these high surface temperatures, according to 1D radiative-convective atmosphere models (Marcq, 2012; Marcq et al., 2017; Pluriel et al., 2018), it was found that the steam atmosphere heights are increasing up to a few thousands of kilometers (Benedikt et al., 2019). The main process, which is dragging heavier atoms into space, is the EUV-driven hydrodynamic escape of H atoms from these extended upper atmospheres, originating from dissociation of $H_2O$ at the 1 µbar (Guo, 2019) level.

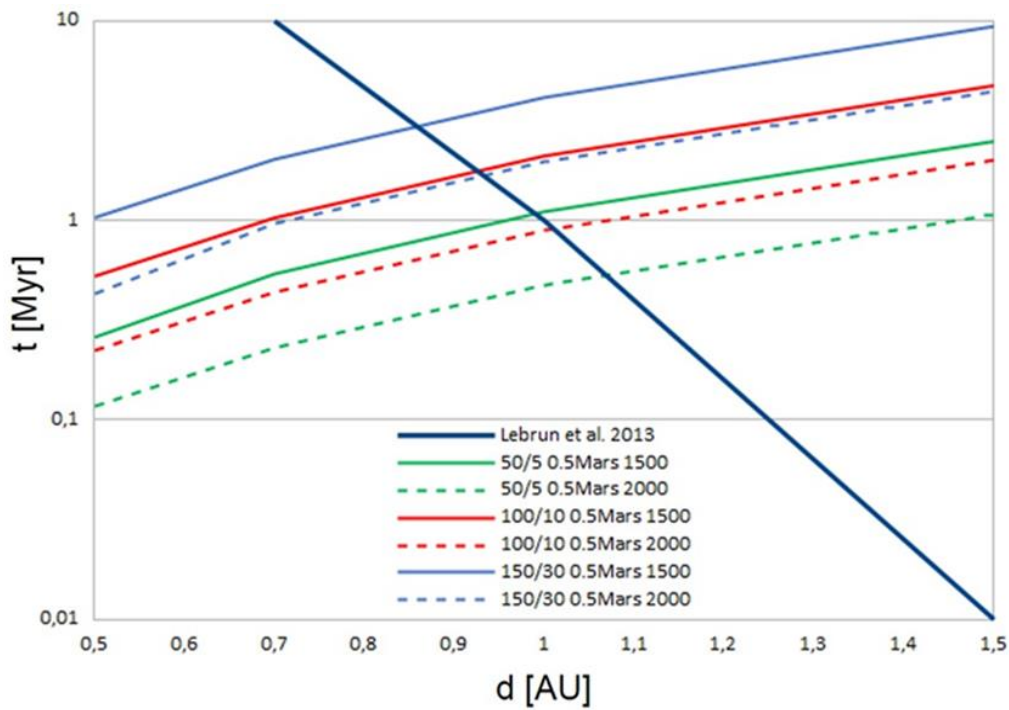

**Fig. 7.** Escape time *t* of steam atmospheres for a planetary embryo with a mass of $0.5 M_{Mars}$, various partial surface pressures of $H_2O/CO_2$ and surface temperatures of 1500 and 2000 K for a weakly active slowly rotating young G-type star in dependence of the orbital distance *d*. The solid dark blue line in each panel shows the steam atmosphere solidification time based on the study of Lebrun et al. (2013) (after Benedikt et al., 2019).

For planetary embryos with masses $\leq 1\,M_{\text{Moon}}$ no dense silicate atmosphere or steam atmosphere can build up (Hin et al., 2017; Benedikt et al., 2010; Young et al., 2019), the embryo will continuously lose outgassed elements very fast to space. Therefore, one can expect that these bodies will be strongly depleted in volatiles, noble gases and outgassed rock-forming elements.

One should note that in the recent study by Young et al. (2019) it is assumed that hydrostatic conditions for an outgassed rock vapour atmosphere above a magma ocean with a surface temperature of 1830 K on a planetary embryo with half of the mass of Pluto builds up. As discussed in Benedikt et al. (2019), under such extreme conditions where the ratio between the gravitational energy and the thermal energy (Jeans parameter) is < 6 (in the case of Young et al., 2019 the Jeans parameter at the surface is 1.38) it is impossible that an atmosphere can build up at such low mass and hot bodies. In such cases, direct escape of outgassed rock-forming elements occurs unless there is a steady supply from the interior to compensate for these fast losses. For silicate atmospheres, this means that the entire mantle will eventually depleted in the outgassed elements. The assumption of a hydrostatic equilibrium as assumed in the study of Young et al. (2019) is thus only justified for cases in which the Jeans parameter of the outgassed elements at the surface of an embryo exceed $\approx 6$ (Erkaev et al., 2015; Volkov et al., 2011).

According to Benedikt et al. (2019) more massive embryos with masses larger than 0.5 Mars masses also lose outgassed heavy trace elements when they are incorporated into solidified magma ocean-related steam atmospheres. As long as steam atmospheres do not condense and outgassed rock-forming elements reach the upper atmosphere they can also be lost by the dragging of escaping H atoms. Fig. 7 shows an example of escape times of magma ocean-related catastrophically outgassed steam atmospheres and embedded trace elements for a planetary embryo with a mass of $0.5 M_{\text{Mars}}$ at various orbital distances, partial surface pressures of 50/5 (green), 100/10 (red), 150/30 (blue) bar $H_2O/CO_2$ and surface temperatures of 1500 K (solid line) and 2000 K (dashed line). The bold blue line corresponds to the condensation temperature of a steam atmosphere based on the results of Lebrun et al. (2013). As one can see in Fig. 7, depending on the partial pressure and surface temperature, embryos with a mass of $0.5 M_{\text{Mars}}$ can lose steam atmospheres at an orbital distance of 1 AU within the condensation time. For heavier planetary embryos of up to $\approx 1.5 M_{\text{Mars}}$ almost all outgassed trace elements can be lost together with the escaping steam atmosphere within $\approx 10$ Myr (Benedikt et al., 2019), which lies within the time frame of the formation of the first protocrust on Mars (Bouvier et al., 2018).

This indicates that frequent smaller impactors (Maindl et al., 2015) keep shallow magma oceans molten most likely over longer time frames so that the steam atmosphere can longer remain in vapour form compared to the modelled time scales given in Lebrun et al. (2013). A more active young Sun-like star leads to a shorter escape time for the whole steam atmosphere compared to the slow rotator as shown in Fig 7.

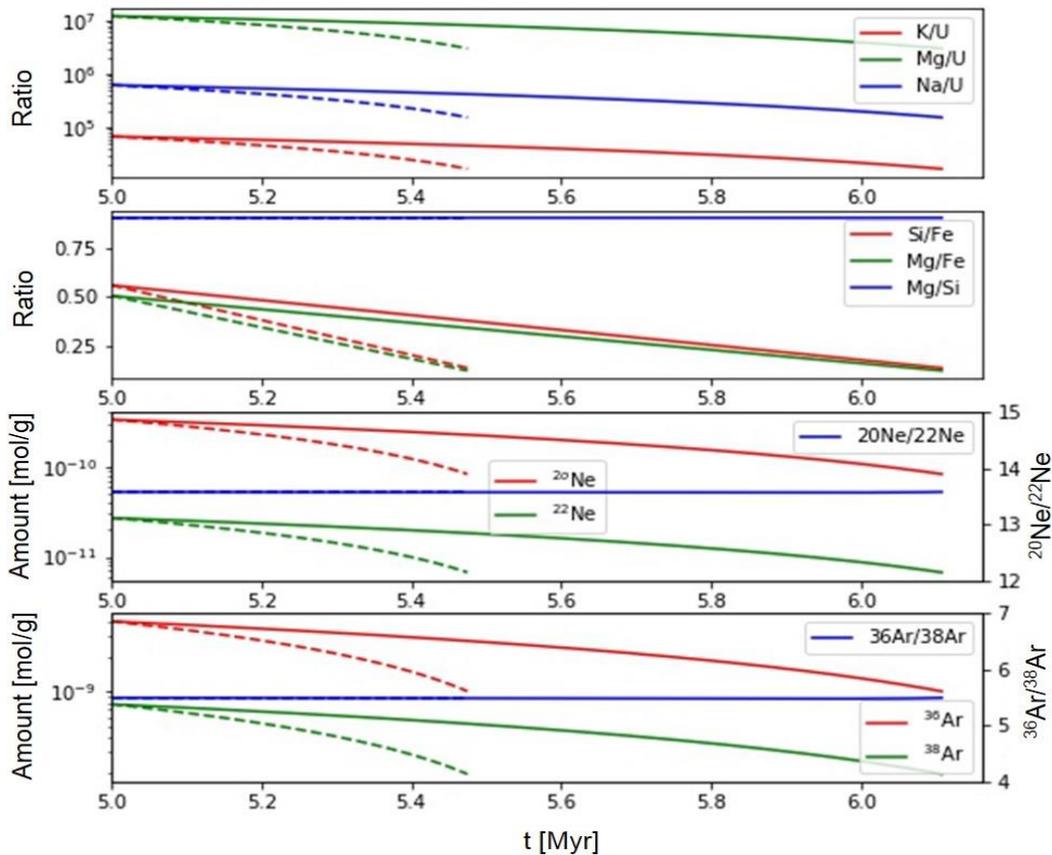

**Fig. 8.** Escape related evolution of the K/U, Mg/U, Na/U, Si/Fe, Mg/Fe, Mg/Si, $^{20}Ne/^{22}Ne$, $^{36}Ar/^{38}Ar$ ratios and losses of Ar and Ne as a function of time of a 50 bar $H_2O$ and 5 bar $CO_2$ outgassed steam atmosphere from a planetary embryo with $0.5 M_{Mars}$ at 1 AU. The embryo's steam atmosphere is exposed to the EUV flux of a weakly active slowly rotating young Sun-like star after the magma ocean solidification that is assumed to be at $\approx 5$ Myr. About 75% of the total initial amounts of the elements are lost more or less within 1 Myr with surface temperatures corresponding to 1500 K (solid lines) and 2000 K (dashed lines). The lines end when the steam atmospheres are lost from the embryo. The ratios were calculated for mol/g and U and Fe assumed not to escape but remain at the body with their initial amounts (after Benedikt et al., 2019).

Fig. 8 shows an example of the remaining amount of various elements on a planetary embryo with 0.5 $M_{Mars}$ at 1 AU, with a magma ocean-related steam atmosphere partial surface pressure of 50 bar $H_2O$ and 5 bar $CO_2$. If all outgassed elements escape, the particular volatiles are depleted by ≈ 75 % within the planetary embryo. The remaining amount of Ne and Ar isotopes in mol/g and $^{20}$Ne/$^{22}$Ne and $^{36}$Ar/$^{38}$Ar ratios are also shown. One can see that on such low mass bodies the Ne and Ar isotopes only escape to space but do not fractionate due to the high loss rates. Isotope fractionation, however, between the vapour and melt of small planetary embryos could at least fractionate the isotopic ratios of Mg and Si at near-equilibrium (Young et al., 2019).

It should be noted, however, that the depletions and ratios of the volatile rock-forming elements of the larger planetary embryos with masses ≥ 0.5 $M_{Mars}$ most likely represent upper values. First, it is unclear whether 100 % of the outgassed elements will indeed be incorporated into the steam atmosphere or if a fraction remains inside the magma ocean. Second, the outgassed rock-forming elements have to overcome a bottleneck from the lower atmosphere to the upper atmosphere where they have to overcome a so-called cold-trap in the mesosphere; otherwise, they will reach the condensation temperature.

For noble gases such as Ne or Ar, however, the atmospheric cold-trap of larger bodies does not exist due to their low condensation temperatures. Because of this, it seems reasonable that large planetary embryos should be more depleted in noble gases (i.e., Ne and Ar) than in rock-forming elements such as K, Si, Mg, Fe, Ca, Al, Na, S, P, Cl, F due to their significantly higher condensation temperatures.

In general, these studies also agree with the recent experimental and theoretical based results of Sossi and Fegley (2018) and Sossi et al. (2019), that atmospheric escape caused by one large evaporation event at Earth is ineffective in fractionating moderately volatile element abundances to an extend that match the observations. Instead, the present moderately volatile element abundance in terrestrial planets most likely reflects accretion from several lower-mass embryos where each of them experienced different degrees of volatilization. One can conclude from these recent studies that a late accretion of chondritic material via planetary embryos that underwent melting processes might therefore not necessarily lead to a volatile rich planet.

# 7 Evolution of Earth's and Venus' Ar and Ne noble gas isotope and K/U bulk elemental ratios

Chondritic meteorites have long been linked to the bulk composition of the silicate part of Solar System terrestrial planets (e.g. Taylor, 1964; Anders and Grevesse, 1989; Javoy et al. 2010; Dauphas 2017; Schiller et al., 2018). Dauphas (2017) analyzed isotopes of lithophile (O, Ca, Ti, Nd), highly siderophile (Ru) and moderately siderophile (Cr, Ni and Mo) elements from Earth's mantle and found that they record different accretion stages. According to analysed isotope compositions by Dauphas (2017), the first 60 % of proto-Earth consisted of about 50 % enstatite-like and 5 % chondrite-like material, whereas all of the remaining accreted material consisted of enstatite-like impactors. According to Budde et al. (2019) most isotopic compositions can also be explained if mixtures of carbonaceous chondritic and non-carbonaceous chondritic material composed the building blocks of the accreting Earth. On the other hand, the oxygen isotopic composition of the Earth is nearly identical to that of enstatite chondrite meteorites. Oxygen is the most abundant element in rocks and one of the most abundant element in Earth depending on the estimated FeO content in the mantle and the Fe content in the core. Unless the Earth has a large enstatite chondrite component, the oxygen-isotopic composition and the iron content of the Earth cannot be matched (Dauphas, 2017). The Dauphas (2017) O-isotope model and composition hypothesis agrees also with the two-component models of Ringwood (1979), Wänke and Dreibus (1988) and Lodders (2000), which assume almost the same percentage of O-isotope related enstatite chondritic material.

Recently, Schiller et al. (2018) investigated the $^{48}$Ca/$^{44}$Ca isotope evolution of the composition of different Solar System objects such as Lunar and Martian meteorites, ordinary and carbonaceous chondrites, angrites, eucrites, ureilites, as well as very young chondrules from ordinary and carbonaceous chondrites and its role as a precursor element to rocky planets. The results of this study indicate that the $^{48}$Ca/$^{44}$Ca composition of the earliest material in the inner disk originated from a ureilite-like composition and evolved to a terrestrial composition within the first million years after the formation of the Sun. Calculated accretion rates of the inner protoplanetary disk inferred from the Ca isotope composition of the involved objects reveal that the bulk disk composition at ≈ 5-6 Myr after the formation of the Solar System consisted of a mixture of about 60 % ureilite-like and 40 % outer Solar System CI chondritic material (Schiller et al., 2018).

In contrast to this, $^{36}$Ar/$^{38}$Ar and $^{20}$Ne/$^{22}$Ne noble gases of the present Earth atmosphere and isotopes of water (D/H), and nitrogen ($^{14}$N/$^{15}$N) can also be reproduced by assuming a volatile content equivalent to a mixture of about 98 % non-chondritic dry material with a contribution of about 2 % undepleted carbonaceous CI-CM-like material (Marty, 2012). Although, the results for the proto-Earth obtained by Dauphas (2017) and Schiller et al. (2018) might seem to contradict the findings of Marty (2012) representing present Earth, it is not necessarily the case. As shown and discussed in Odert et al. (2018), Benedikt et al. (2019) and in Sect. 6, the building blocks of terrestrial planets were most likely depleted in noble gases and other moderately volatile rock-forming elements. Additionally one should not forget that proto-Earth and Venus underwent different evolutionary processes since the evaporation of the disk, including the possible escape of a primordial atmosphere, magma oceans and impactors with a depletion-modified abundance of their composition. These processes certainly played a crucial role in shaping the noble gas isotope, elemental and volatile evolution of these planets over their history.

Since decades, various researchers tried to model the compositional evolution of early Venus and Earth by reproducing the observed isotope ratios of the non-radiogenic noble gases Ar and Ne in their atmospheres (e.g., Cameron, 1983; Pepin, 1991; 1997; Becker et al., 2003; Gillmann et al., 2009). For instance, Gillmann et al. (2009) studied the escape and fractionation of $^{36}$Ar/$^{38}$Ar, and $^{20}$Ne/$^{22}$Ne noble gas isotope ratios in Venus early atmosphere with a simple hydrodynamic escape model that is based on approaches and inputs provided by earlier studies of Hunten (1997), Kasting and Pollack (1983), Zahnle, and Kasting (1986) and Chassefière (1996a; 1996b; 1997).

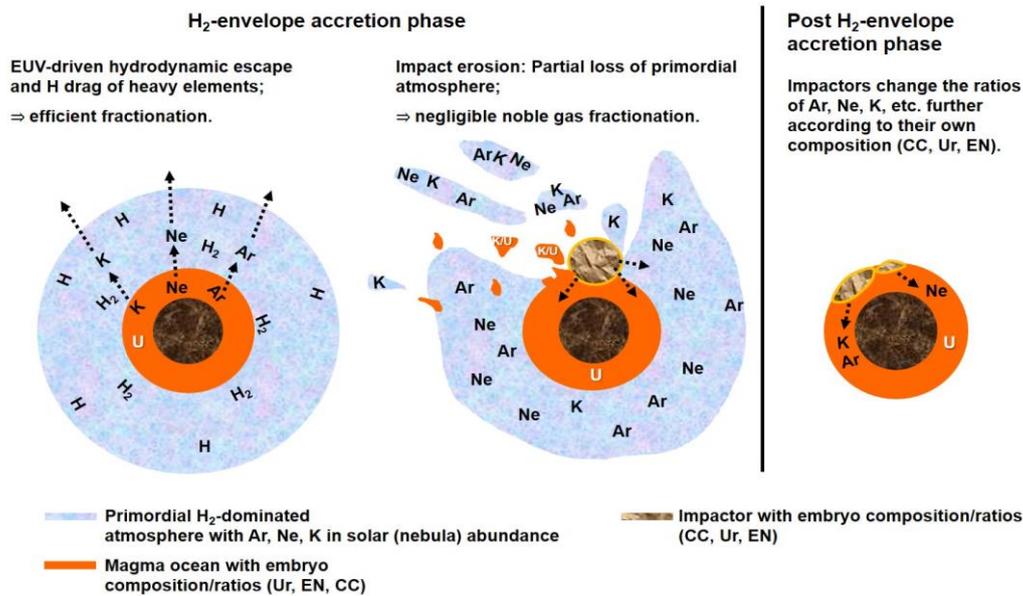

**Fig. 10.** Illustration of main processes and scenarios involved in the reproduction attempts of present day atmospheric 36Ar, 38Ar, 20Ne, 22Ne, noble gas isotope ratios and bulk K/U ratios on Venus and Earth by Lammer et al. (2019).

In this approach Gillmann et al. (2009) introduced a factor $\eta = \varepsilon f + \Phi_{sw} / \Phi_{EUV}$, that depends on the heating efficiency $\varepsilon$ of the solar EUV flux $\Phi_{EUV}$, generally assumed to be ≈ 15 % (Chassefière, 1996b; Shematovich et al., 2014; Lammer et al., 2014), a geometrical amplification factor $f$ related to the EUV heated and extended upper atmosphere and the relative contribution $\Phi_{sw}$ of the solar wind energy deposition to the total energy deposition (solar wind + EUV) near the exobase level. Since at the time of this study only very rough estimates existed on the young Sun's activity evolution, Gillmann et al. (2009) assumed $\eta$ values between 0.3 – 20. The introduced geometrical amplification factor $f$ was also based on arbitrary assumptions within a range of 4 – 8, since no upper atmosphere structures have been modelled in detail in these pioneering but simplified models.

Fig. 9 shows resulting $^{36}Ar/^{38}Ar$ and $^{20}Ne/^{22}Ne$ isotope fractionation patterns in Venus atmosphere, obtained by the simple model approaches presented in Gillmann et al. (2009). One can see that their obtained evolutionary scenarios are consistent within the present error bar ranges. Since no upper atmosphere structure and temperature profiles were modelled in their approach these authors assumed various average thermospheric temperatures between 300 and 2000 K. For high thermospheric temperatures > 1000 K, Gillman et al. (2009) obtained too large isotopic fractionations, because the densities of Ne and Ar and their escape fluxes are too large in their model approach. One can also see that for average thermosphere temperatures < 400 K, Ne is not fractionated enough to fit the present fractionation value due to its low abundance at the escape level. One should also note that the amount of hydrogen necessary to obtain the fractionation results for assumed $\eta$ values of 0.6 (Fig. 9a) and 1.2 (Fig. 9b) contained 6 – 14 terrestrial oceans (TO) and 30 – 40 TOs, respectively (Gillmann et al., 2009).

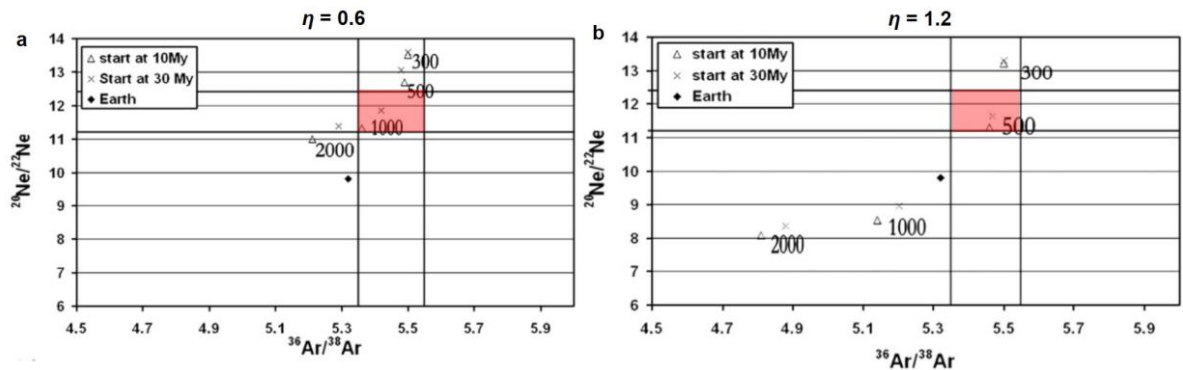

**Fig. 9.** Reproduction attempts of $^{36}Ar/^{38}Ar$ and $^{20}Ne/^{22}Ne$ isotope ratios according to different evolution scenarios related to atmospheric escape of early Venus with assumed average thermospheric temperatures between 300 – 2000 K. The two horizontal and vertical lines, limit the ranges of the measured values of the isotopic ratios of Ne and Ar. They are delineated with the shaded area, which marks the scenarios that are in agreement with the measured isotopic ratios are satisfied. a.) Assumed initial atmospheric hydrogen content based on $H_2O$-amounts of 4 - 14 terrestrial oceans (TOs) for a $\eta$ value of 0.6. b.) Assumed initial atmospheric hydrogen content based on $H_2O$-amounts of 20 - 40 TOs for a $\eta$ value of 1.2. Atomic oxygen does not escape in these scenarios (Gillmann et al., 2009).

More recently, Lichtenegger et al. (2016) studied the expected relative contribution $\Phi_{sw}$ of the solar wind energy deposition on early Venus, by assuming that the young Sun was either a weak or a moderately active young G star. These authors assumed a hydrogen-dominated upper atmosphere and modelled the precipitation of energetic hydrogen atoms (ENAs) that originate due to charge exchange between the planet's extended exosphere and solar wind protons (Chassefière, 1996b; 1997). Chassefière (1997) and Gillmann et al. (2009) expected that these ENAs deposit a huge amount of energy into the upper atmosphere and as a result enhance the H escape rate that influenced the noble gas fractionation. According to Lichtenegger et al. (2016), however, ENAs deposit their energy and heat mainly above the main EUV energy deposition layer. Although these particles modify the thermal structure of the upper atmosphere, the enhancement of the thermal escape rates caused by ENAs remains negligible (Lichtenegger et al., 2016). Since several pioneering studies including that of Cameron (1983), Pepin (1991) and Gillmann et al. (2009) lacked knowledge on the X-ray and EUV evolution of the young Sun/stars, magma ocean physics or impact studies, assumed arbitrarily reservoirs of $H_2$-molecules from primordial atmospheres and/or catastrophically outgassed $H_2O$ inventories, one can consider that these arbitrarily chosen assumptions are insufficiently accurate and hence outdated.

Recently, Odert et al. (2018) tried to reproduce the $^{36}Ar/^{38}Ar$ and $^{20}Ne/^{22}Ne$ isotope ratios of Venus' atmosphere within a more realistic parameter space and found that EUV-driven hydrodynamic escape of an $H_2$-envelope that remained around proto-Venus after the disk evaporated can reproduce the observed atmospheric Ne and Ar isotope ratios within their large error bars. In the study of these authors the EUV activity evolution of the young Sun corresponded to a slowly to moderately rotating young Sun (Tu et al., 2015; Lammer et al., 2019b). These findings stimulated Lammer et al. (2019b) to apply more complex models to figure out whether one can reproduce the measured atmospheric $^{36}Ar/^{38}Ar$, $^{20}Ne/^{22}Ne$, $^{36}Ar/^{22}Ne$ noble gas isotope ratios and bulk planet K/U elemental ratios of Earth and Venus to constrain their early evolution.

That a fraction of K compared to its initial composition was lost during accretion can be seen in the K/U ratio of each planet (e.g. ≈ 13 800 for the bulk silicate Earth (BSE) (Arevalo et al., 2009; Lodders et al., 2009), and ≈ 7 000 on Venus' surface (Davis et al., 2005), compared to a K/U abundance ratio of ≈ 64 650 in the solar photosphere (Lodders et al., 2009) and ≈ 68 930 in unprocessed chondrites (Lodders et al., 2009). This study, for the first time, additionally included the moderately volatile rock-forming element K which is released at high temperatures (≈ 1500 – 2500 K) from magma oceans that form below primordial atmospheres (Schaefer and Fegley, 2010; Fegley et al., 2016; see also O'Neill et al., this issue 2019).

The model efforts carried out in Lammer et al. (2019b) are illustrated in Fig. 10 and summarized below:

- Protoplanetary masses between 0.5 – 0.8 $M_{Earth}$ are placed within the disk at Venus and Earth orbit locations. In case a primordial $H_2$-envelope survives boil-off during

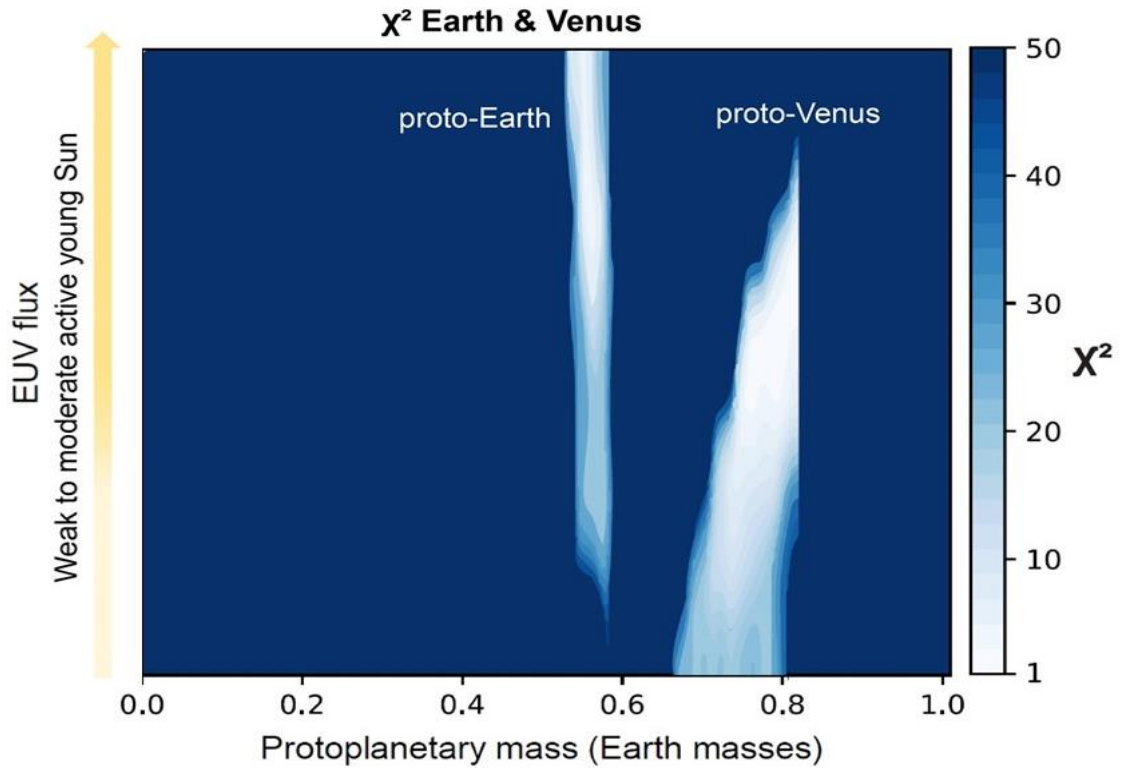

**Fig. 11.** Statistical $\chi^2$ parameter study for constraining proto-planetary masses of early Earth and Venus at the time when the disk evaporated at about 3-5 Myr after the origin of the Solar System by successful reproduction (light areas) of today's atmospheric $^{36}Ar/^{38}Ar$, $^{20}Ne/^{22}Ne$ and bulk K/U ratios. $\chi^2$ values that are larger than 5.0 indicate that at least one element is not within the error bars of today's measurements and scenarios with values larger than 30 show distinct deviations (after Lammer et al., 2019).

- Hydrodynamic upper atmosphere models (Lammer et al., 2014; Kubyskina et al., 2018a; 2018b) that can calculate the EUV-driven escape of H atoms and an evolution code that drag heavier trace elements (Zahnle, and Kasting, 1986; Odert et al., 2018) out of the gravity field of accreting proto-planets are applied for various expected EUV-flux scenarios from slow to moderate and fast rotating young Sun evolution scenarios.

- Atmospheric mass loss of the $H_2$-envelope caused by Moon- to Mars-mass impactors (Sect. 5) is considered additionally to the EUV-driven thermal escape. Since the masses of these impactors, which are assumed to be admixtures of CC, UR, EN, are added to the protoplanetary mass, they are also altering the $^{36}Ar$, $^{38}Ar$, $^{20}Ne$, $^{22}Ne$, and K content of proto-Earth and Venus, which has to be taken into account.

- Evolution of the photospheric radii according to these mass additions, gravity change and atmospheric escape as well as its effects on the isotope/element ratio evolution is also included in the study.

The illustrations in Fig. 10 show the two main phases and processes that are included in the recent model simulation attempts of Lammer et al. (2019b) for reproducing proto-Venus' and Earth's $^{36}$Ar, $^{38}$Ar, $^{20}$Ne, $^{22}$Ne, K/U ratios. Fig. 11 shows the statistical outcome of modelled proto-Earth and proto-Venus masses at the end of the disk lifetime (3 – 5 Myr) by Lammer et al. (2019). The white areas illustrate the parameter space of EUV flux and protoplanetary mass in which the $^{36}$Ar/$^{38}$Ar and $^{20}$Ne/$^{22}$Ne noble gas isotope ratios and bulk planet K/U elemental ratios were successfully reproduced for particular compositions on both planets within their error bars.

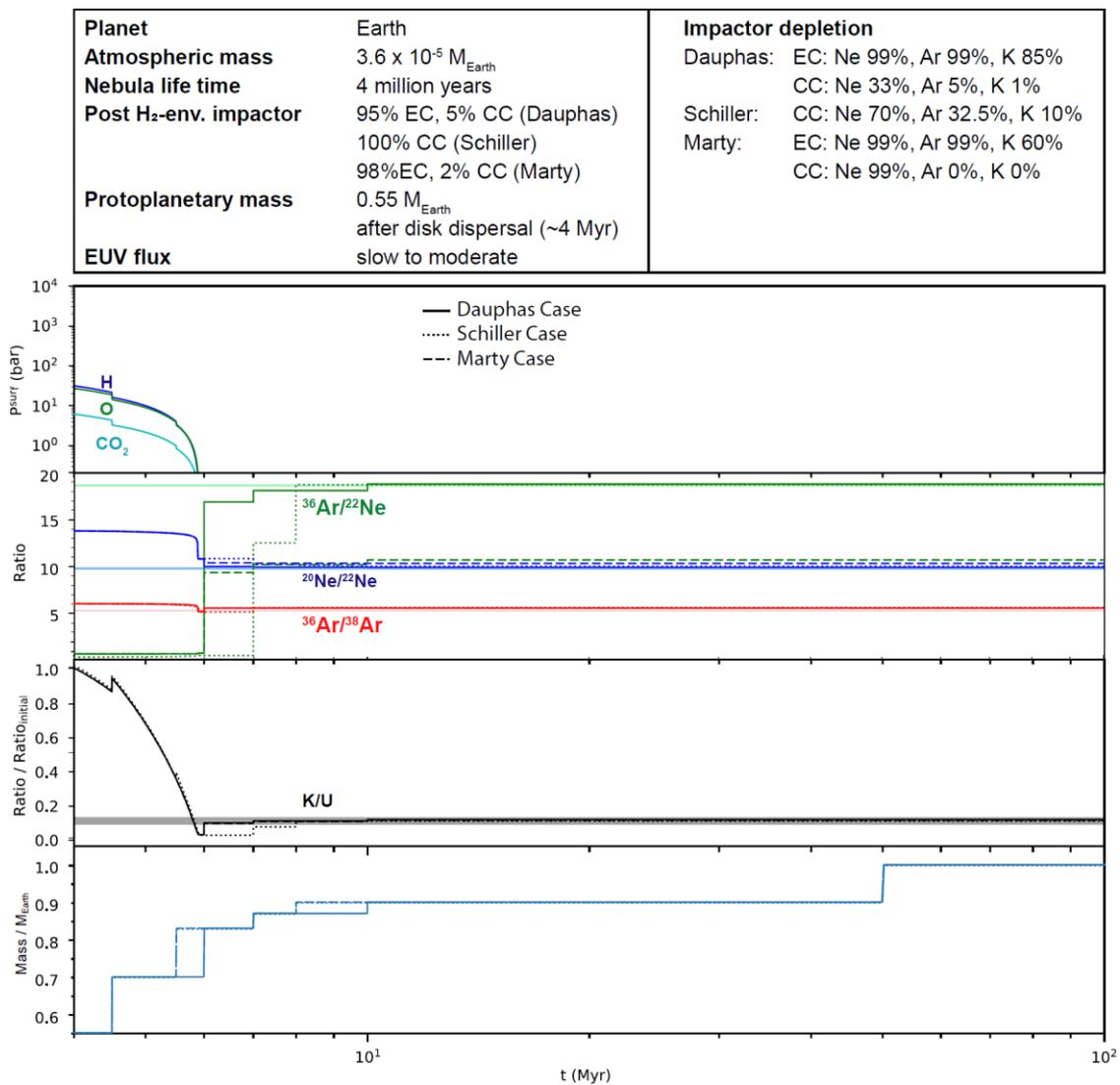

**Fig. 12.** Reproduction attempts of Earth's present-day atmospheric $^{36}$Ar/$^{38}$Ar, $^{20}$Ne/$^{22}$Ne, $^{36}$Ar/$^{22}$Ne and bulk K/U ratios by assuming different post-H$_2$-envelope impactor compositions.

One can see that the present day ratios can be reproduced if proto-Earth and Venus were released from the disk between 3 – 5 Myr with masses of about 0.53 – 0.58$M_{Earth}$ and 0.68 – 0.81$M_{Earth}$, respectively, if the young Sun followed an EUV flux evolution path between a slowly and moderately rotating young G-type star.

Fig. 12 shows evolution scenarios for early Earth with an initial protoplanetary mass of 0.55 $M_{Earth}$ at the end of the disk at ≈ 4 Myr (Bollard et al., 2017; Wang et al., 2017) orbiting a young Sun that evolves along an EUV activity that corresponds to an intermediate rotating young G-type star, with a post-$H_2$-envelope impactor composition of ≈ 95 % EN and ≈ 5 % CC (Dauphas, 2017; solid lines), 100 % CC (Schiller et al., 2018; dotted lines) and 98 % EC and 2% CC (Marty, 2012; dashed lines). Today's atmospheric $^{36}Ar/^{22}Ne$ ratio of ≈ 18.8 could be best reproduced by the model methods applied in Lammer et al. (2019) for the Dauphas and Schiller scenarios but with different assumed growth rates and depletions of the post-$H_2$-envelope impactors. A significant difference in these production attempts is the amount of volatile rich CCs. The more CCs are delivered to the growing protoplanet the faster it has to grow so that the volatile elements can escape together with the captured primordial $H_2$-dominated atmosphere. If proto-Earth captured such a tiny primordial atmosphere, initial composition suggested by Marty (2012) cannot reproduce Earth's present atmospheric $^{36}Ar/^{22}Ne$ ratio. However, if proto-Earth accreted less mass so that no primordial atmosphere remained after disk dispersal the composition derived in Marty (2012) cannot be excluded. Due to this, the reproduction of the present day atmospheric Ar and Ne isotope ratios only yields an upper limit for the mass of the proto-Earth (≤ 0.58 $M_{Earth}$) at the time when the disk evaporated.

According to the resent work of Burkhardt et al. (2019) an analysis of measured Ti and Sr isotopic compositions of Ca,Al-rich inclusions (CAIs) from the Allende CV3 chondrite indicate that CAI-like material was the carrier of the nucleosynthetic isotopic anomalies in most minor and trace elements. These authors found that a variable admixing of CAI-like refractory material to an average inner solar nebula component could explain the planetaryscale Ti and Sr isotope anomalies and the elemental and isotopic difference between noncarbonaceous (NC) and carbonaceous (CC) nebular reservoirs for these elements. If this is the case, then the differences between meteorites and Earth may simply reflect the timing or spatial location for adding this CAI-like material to the solar nebula. The findings by Lammer et al. (2019) that the post-$H_2$-envelope impactors should have also be at least partially related to the CC-reservoir agrees also with the dating of terrestrial planet building blocks from CAIs, from which it is inferred that CCs finished their accretion at times around ≈ 3 – 4 Myr after the formation of the Solar System (Kruijer et al, 2017).

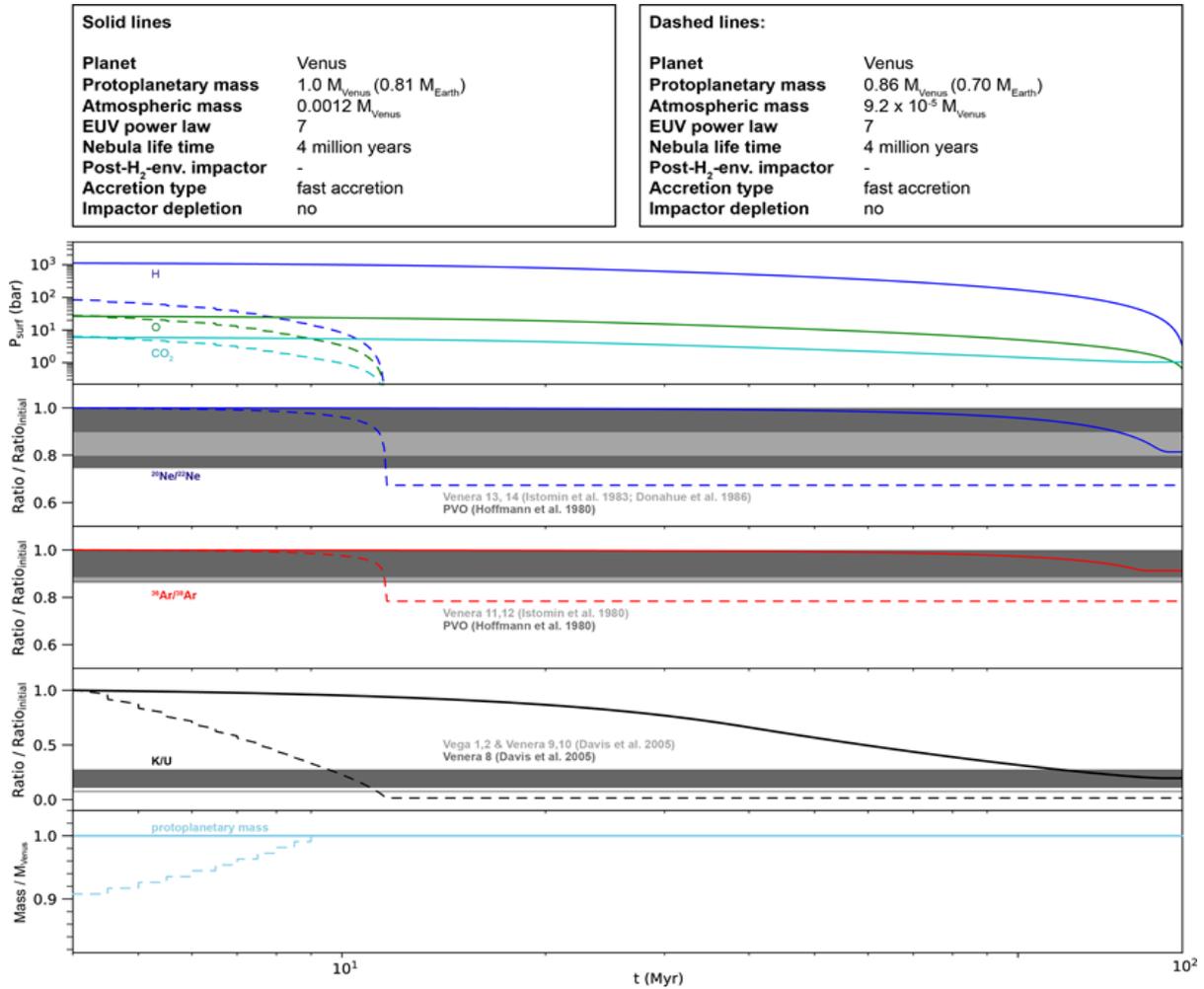

**Fig. 13.** Successful (solid lines) and failed (dashed-lines) reproduction attempts of the atmospheric $^{36}Ar/^{38}Ar$ and $^{20}Ne/^{22}Ne$ isotope and K/U surface ratios at Venus. A final accreted Venus at the time when the disk evaporated can reproduce the present observations if the captured primordial $H_2$-dominated atmosphere is exposed to a slowly to less than moderately rotating young Sun (solid lines). All ratios are normalized to the initial ratio. A lower mass proto-Venus with a mass of about $0.86 M_{Venus}$ over fractionates these noble gas isotopes and yields lower K/U ratios as measured by Venera 8, 9, 10, and Vega 1 and 2 (after Lammer et al., 2019).

Fig. 13 shows a best-case evolution scenario obtained by Lammer et al. (2019) that can reproduce the present atmospheric $^{36}Ar/^{38}Ar$, $^{20}Ne/^{22}Ne$ and measured surface K/U ratios on Venus by assuming a similar initial composition and the same young Sun EUV flux evolution track as assumed for the successful reproduction of the elements on Earth in Fig. 12. In this particular scenario, proto-Venus accreted its whole mass during the disk lifetime and was released from the disk at ≈ 4 Myr with a $H_2$-dominated primordial atmosphere with a hydrogen partial pressure of ≈ 1000 bar. The escaping H atoms dragged away the heavier trace species and fractionated $^{36}Ar/^{38}Ar$, $^{20}Ne/^{22}Ne$ from the initially solar-dominated ratios to the ones in today's Venus' atmosphere (Lammer et al., 2019). Besides the noble gas isotope ratios, the corresponding K/U ratio lies slightly above Vega 1, 2 and Venera 9, 10 data (Davis et al., 2005) but within the range of the Venera 8 landing site (Basilevsky, 1997; Abdrakhimov and Basilevsky, 2002). The dashed lines in Fig. 12 show an example modelled by Lammer et al. (2019) where these authors failed to reproduce the present-day ratios for Venus. In this failed attempt a less massive proto-Venus with about 0.86 $M_{Venus}$ (0.69 $M_{Earth}$) leads to an under fractionation of the Ar, Ne isotope and K/U ratios.

Compared to both large terrestrial planets in the Solar System, the evolution for the lower mass planet Mars was different and one cannot constrain it in a similar way as in the case of Venus and Earth. Mars' mass is too low to capture a relevant $H_2$-envelope during the phase the small planet grew within the gas disk. Noble gases could have also been lost and dragged away by the escaping H atoms as long as hydrogen, which originated from solidified magma ocean-related outgassed steam atmospheres, dominated the upper atmosphere (Pepin, 1994; 2002). Afterwards ionization related fractionation through thermal and non-thermal escape (i.e., ion pick up, sputtering, dissociative recombination) of the planet's $CO_2$ atmosphere resulted in today's observed Ar, Ne and N isotopic compositions (e.g., Jakosky et al., 1994; Hutchins and Jakosky, 1996; Carr, 1999). In the next section, the isotopic fractionation and escape of the Martian atmosphere and initial water inventory are discussed.

# 8 Isotopic fractionation as a constraint on the atmospheric evolution and water inventory on Mars

Small gravity and absence of an intrinsic magnetic field allow efficient atmospheric escape from Mars induced by the solar EUV radiation and the solar wind. Its isotopic compositions show heavy-isotope enrichment in hydrogen (D/H), carbon ($^{13}C/^{12}C$), nitrogen ($^{15}N/^{14}N$), and noble gases ($^{22}Ne/^{20}Ne$, $^{38}Ar/^{36}Ar$, and non-radiogenic xenon isotopes) except krypton compared to their possible origins (nebula gas, chondrites, and comets), as an outcome of atmospheric escape. As mentioned above, noble gases are particularly useful to constrain the atmospheric evolution since they are mostly partitioned into the atmosphere and do not form molecules so that the escape processes are limited to hydrodynamic escape and to atmospheric sputtering for lower mass bodies without magnetospheres (e.g., Pepin, 1991; 1994; Jakosky et al., 1994). Nitrogen behaves similarly to noble gases as it is chiefly partitioned into the atmosphere but ion-molecule photochemical reactions also induce atmospheric escape (e.g., Fox and Dalgarno, 1983; Fox 1993ab; Fox and Hác, 1999; see also Avice and Marty 2019; this issue). Carbon is partitioned into multiple reservoirs - the atmosphere, carbonate, organic compounds, $CO_2/CH_4$-clathrate, and $CO_2$ ice (e.g., Kurahashi-Nakamura and Tajika, 2006; Longhi, 2006; Hu et al., 2015; Wray et al., 2016) so that the isotopic fractionation is more complicated.

As briefly discussed in Sect. 4 fractionation mechanisms include diffusive separation by mass between the homopause and exobase (e.g., McElroy and Yung, 1976; Jakosky et al., 1994; 2017). The number densities of photochemically long-lived atmospheric species follow their own scale height above the homopause. As the escape processes remove atoms, molecules, and ions from the vicinity of the exobase, the diffusive separation causes mass-dependent fractionation. Since atmospheric sputtering is so energetic that mass-dependent fractionation in the removal process itself is limited (Johanson, 1992), the isotopic fractionation through sputtering is chiefly determined by the diffusive separation (Jakosky et al., 1994).

The diffusive separation between the homopause and exobase was recently probed with the help of the MAVEN spacecraft (Jakosky et al., 2017; Slipski et al., 2018). Measurements of the $N_2/^{40}Ar$ ratio indicated substantial variations at the homopause ($\approx 60 - 140$ km) and exobase ($\approx 140 - 200$ km) altitudes. This represents an unresolved combination of real geophysical variations with time and location but confirms the concept of elemental and isotopic fractionation. Photochemical processes fractionate isotopic ratios due to the difference in reaction rates between isotopic species. Both photodissociation of CO and dissociative recombination of $CO^+$ and $CO_2^+$ could induce preferential loss of $^{12}C$ rather than $^{13}C$ (Fox and Hác, 1999; Hu et al., 2015). Dissociative recombination of $N_2^+$ causes isotopic fractionation between $^{14}N$ and $^{15}N$ (Wallis, 1989; Fox and Hác, 1997; Manning et al., 2008).

### 8.1 Isotopic compositions of the present-day Martian atmosphere

The observed isotopic compositions of $N_2$, Ne, and of the non-radiogenic Ar ($\delta^{15}N = 572 \pm 82$ ‰, $^{20}Ne/^{22}Ne = 10.1 \pm 0.7$, and $^{36}Ar/^{38}Ar = 4.2 \pm 0.1$, Pepin, 1991; Wong et al., 2013; Atreya et al., 2013) are well described by the steady state where the loss via atmospheric escape balances the supply via, for example, volcanic outgassing (Fig. 14), and is given by $I/I_0 = 1/f$, where $I$, $I_0$, and $f$ are the isotope ratio in the atmosphere, the isotope ratio of the source, and the net fractionation factor.

The estimated lifetimes of N (via sputtering and photochemical escape) and Ne (via sputtering) are shorter than one billion years, supporting the hypothesis of steady state (Jakosky et al., 1994; Manning et al., 2008; Kurokawa et al., 2018). The lifetime of Ar is comparable to the age of Mars (Jakosky et al., 1994), but its isotopic ratio being close to the value of steady state suggests that the lifetime was shorter on earlier time (Kurokawa et al., 2018). Being close to the steady state suggests

- the model for their isotopic fractionation is valid, and
- the present-day isotopic compositions do not reflect the earlier period (i.e. before ≈ 4 billion years ago (Ga)).

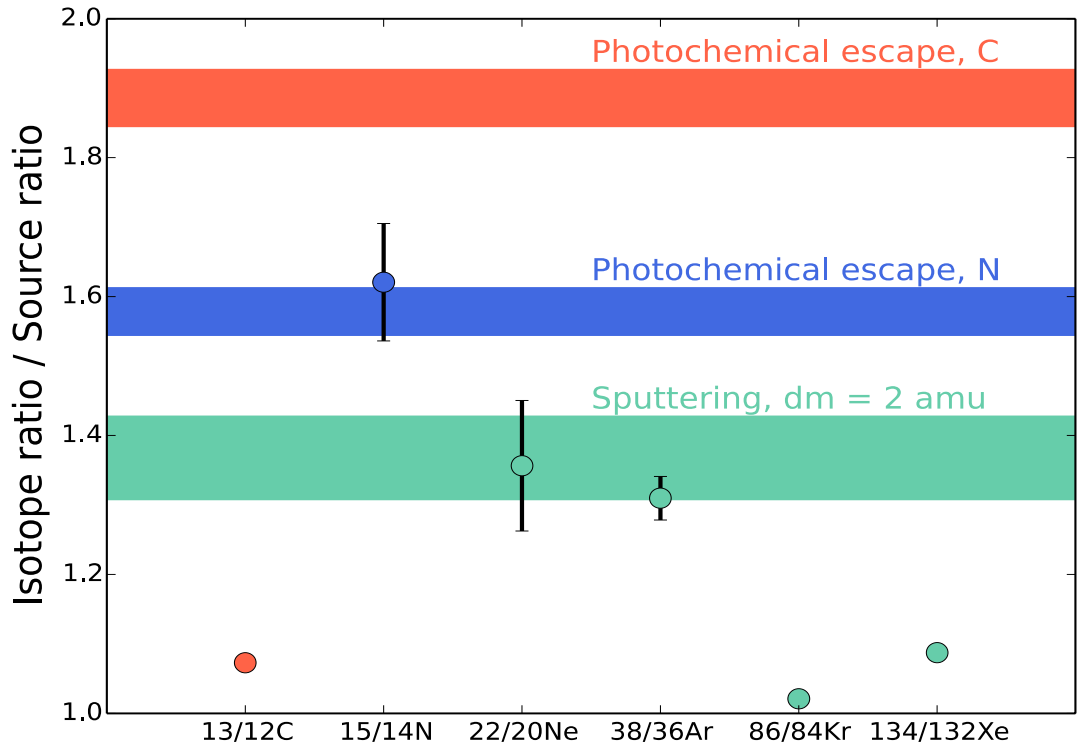

**Fig. 14.** Isotope ratios of volatile elements in present-day Martian atmosphere with respect to their source isotope ratio (data points). We assumed the source values of Martian mantle informed from Martian meteorites for C and N (Wright et al. 1992; Mathew and Marti 2001), and solar abundances for noble gases (Pepin 1991; Pepin et al. 2012). Data for Martian atmosphere are from Mahaffy et al. (2013), Pepin (1991), Wong et al. (2013), Atreya et al. (2013), Conrad et al. (2016). The expected steady state values are shown as horizontal ranges. We assumed the upper atmospheric temperature to be 200 K and the homopause-exobase separation 60-80 km to calculate the fractionation factor for the diffusive separation. We adopted additional fractionation for photochemical escape of C and N, $f = 0.62$ and $0.74$, respectively (Manning et al., 2008; Hu et al., 2015).

The observed $^{13}C/^{12}C$ ratio being smaller than the ratio expected from the steady state with sputtering and/or photochemical escape (Fig. 14) suggests that atmospheric escape is not the only sink of an ancient dense atmosphere. Other sinks include carbonate precipitation (Hu et al., 2015; Wray et al., 2016), organic compounds deposition, and $CO_2$-clathrate formation following the basal melting of $CO_2$-ice caps (Longhi, 2006; Kurahashi-Nakamura and Tajika, 2006). Hu et al. (2015) considered atmospheric escape, carbonate precipitation, and concluded that Mars at ≈ 3.8 Ga had an atmospheric pressure ≤1 bar. Based on an experimental study (Eiler et al., 2000), part of the atmospheric $CO_2$ may convert into $CO_2$ ice without significant isotopic fractionation ($\delta^{13}C_{ice-vapor} <$ -0.4 ‰ at 135 K). Considering additional sinks may allow a denser atmosphere.

The isotopic composition of Xe is also less fractionated compared to the value expected from the steady state with atmospheric sputtering. It suggests that the present-day Xe is a remnant of early hydrodynamic loss (Pepin, 1994) or a result of ionization-induced escape (Zahnle et al., 2019; see also Avice and Marty, 2019; this issue). Ionization-induced escape can lead to high $Xe^+$ escape rates, while the escape of other heavy noble gases such as the hardly ionizable Kr is not significant.

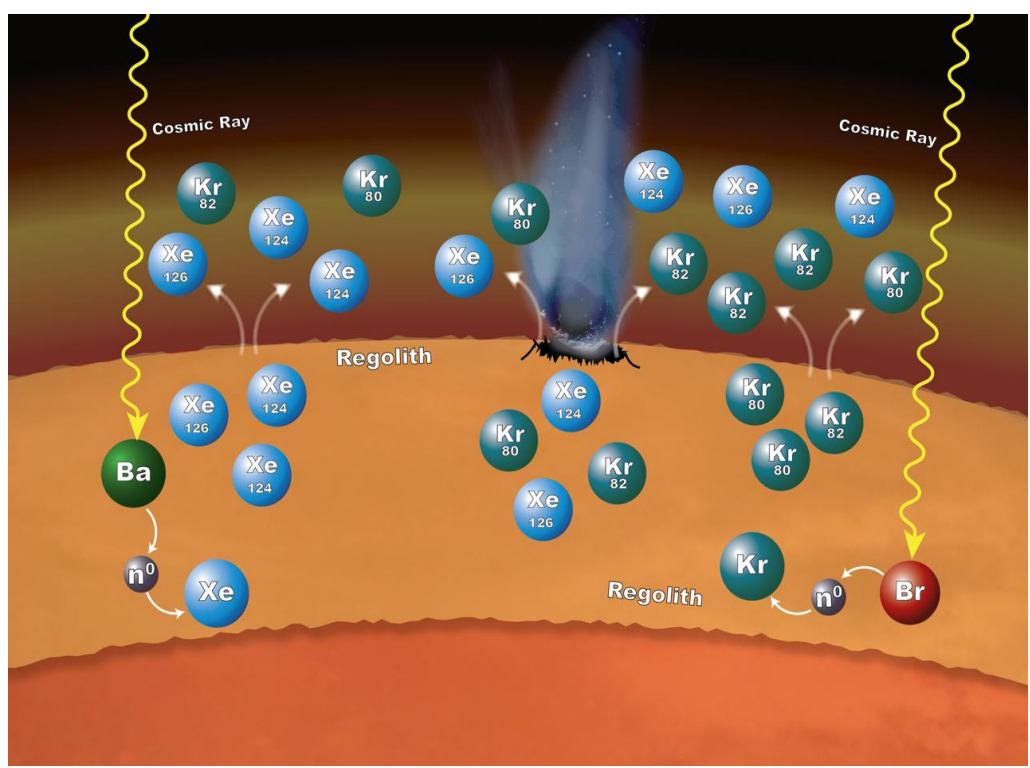

**Fig. 15.** Illustration of the cosmic ray-related surface chemistry that takes place in the Martian regolith, as an explanation why particular Xe and Kr isotopes are more abundant in the atmosphere than expected (Conrad et al., 2016) (NASA/GSFC/JPL-Caltech).

Furthermore, as illustrated in Fig. 15, atmosphere-surface interaction processes in combination with cosmic ray exposure and its related chemistry in the Martian surface can also contribute to an enhancement of some of the Xe and Kr isotopes in the Martian atmosphere (Conrad et al., 2016). Because of the present thin Martian atmosphere, galactic cosmic rays penetrate into the surface material. According to these authors, if cosmic rays hit a barium (Ba) atom, Ba can lose neutrons. Xe atoms can catch some of those neutrons to form the isotopes $^{124}$Xe and $^{126}$Xe.

In a similar way, Br can lose neutrons to Kr, leading to the formation of $^{80}$Kr and $^{82}$Kr isotopes. Impacts and abrasion allow that these isotopes enter the atmosphere, where they underwent escape by impact erosion.

### 8.2 Hydrodynamic escape from early Mars

Models of the upper atmosphere have shown that hydrodynamic escape prohibits early Mars from developing a stable atmosphere (Tian et al., 2009; Erkaev et al., 2014; Odert et al., 2018). The primordial H$_2$/He atmosphere originated from solar nebula (Pepin 1991, 1994; Stökl et al., 2015; 2016; Saito and Kuramoto, 2018) but was lost due to a short and efficient boil-off phase during disk dispersal (Lammer et al., 2018). This was followed by a catastrophically outgassed steam atmosphere during the solidification of a magma ocean (Elkins-Tanton, 2008; Lebrun et al., 2013; Erkaev et al., 2014) that has been lost early on, and a secondary atmosphere later from volcanic outgassing (Lammer et al., 2013).

Hydrodynamic escape involves mass-dependent elemental and isotopic fractionation: heavier species are bound by the gravity more strongly and enrich in the remaining atmosphere (e.g., Hunten et al., 1987; Pepin, 1991; 1994; Odert et al., 2018). The isotopic compositions of trapped non-radiogenic Xe in Martian meteorites Allan Hills 84001 (ALH84001) formed 4.1 Ga and North West Africa 7034 (NWA7034) formed 4.4 Ga are indistinguishable from current atmospheric Xe (Cassata, 2017; see also Avice and Marty; this issue 2019), suggesting that the fractionation has completed before 4.4 Ga. If atmospheric Xe fractionated before 4.4 Ga and was later influenced by impact erosion and replenishment through time (de Niem et al., 2012; Kurokawa et al., 2018; Sakuraba et al., 2019), atmospheric escape of Xe in the later period via a photo-ionized hydrogen wind (Zahnle et al., 2019) might compensate the replenishment.

### 8.3 Atmospheric evolution after the build-up of a secondary atmosphere

Since except Xe the present-day isotopic compositions do not probe the early atmospheric history, Kurokawa et al. (2018) utilized the isotopic compositions in the atmosphere 4.1 Ga recorded in ALH84001 (Mathew and Marti, 2001). While a dense atmosphere preserves primitive isotopic compositions in their model, a thin atmosphere on early Mars is severely influenced by stochastic impact events and following escape-induced fractionation (Fig. 16).

Though the timing when Mars started to build up a stable atmosphere depends on the initial spin rate and related activity of the Sun (Tu et al., 2015) as well as on the volcanic outgassing activity on early Mars (Amerstorfer et al., 2017; Lammer et al., 2018), the less fractionated N ($\delta^{15}N = 7‰$) and Ar ($^{36}Ar/^{38}Ar > 5.0$) recorded in the ALH84001 meteorite (Mathew and Marti, 2001) which formed at 4.1 Ga (Lapen et al., 2010) suggest that a dense atmosphere whose pressure was > 0.5 bar has already existed at the time (Kurokawa et al., 2018).

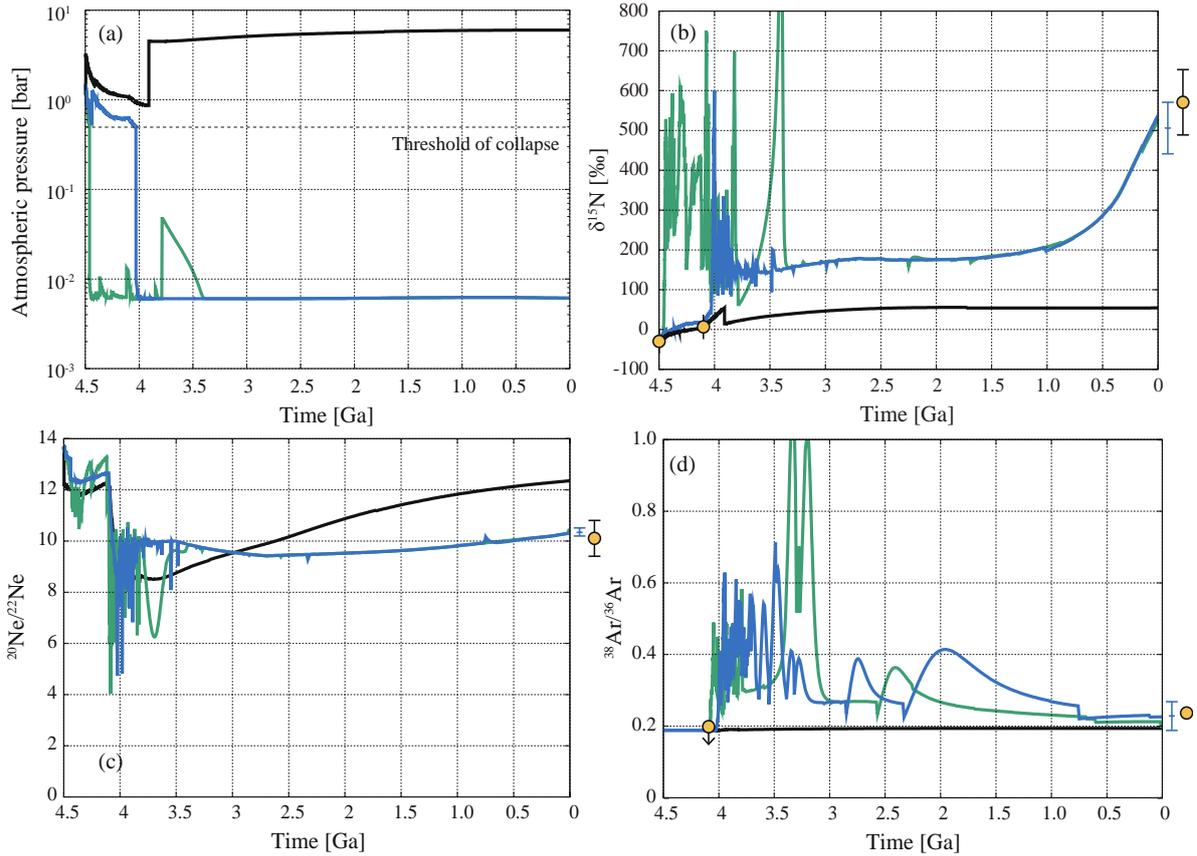

**Fig. 16.** Results of Monte Carlo simulations for the evolution of (a) atmospheric pressure, (b) N, (c) Ne, and (d) Ar isotope ratios (modified after Kurokawa et al., 2018). Three cases are shown: the atmosphere has been dense for 4.5 Gas (black), the atmosphere becomes thin after 4.1 Ga (blue), and the atmosphere becomes thin before 4.1 Ga (green). Data points in blue show the average value in the cases where the atmosphere collapsed at some point. Data points in yellow are the value of Martian atmosphere (Wong et al., 2013; Pepin, 1991; Atreya et al., 2013; Mathew and Marti, 2001).

Other geochemical and geological estimates of paleo-pressure also point to a moderately dense atmosphere whose pressure is comparable to or slightly ≤1 bar (Kite 2019 and references therein). Elevated radiogenic Ar ($^{40}Ar/^{36}Ar$) and Xe ($^{129}Xe/^{132}Xe$) ratios recorded in ALH84001 (Cassata et al., 2010; Cassata 2017) are consistent with the Martian atmosphere originating at ≈ 4.1 Ga from late outgassing.

## 8.4 Loss of water on Mars

The D/H ratio recorded in ALH84001 (Boctor et al., 2003; Greenwood et al., 2008; Usui et al., 2017) is in the range of $\delta D = 500 – 3000$ ‰, which means that more than 1/3 of the surface water could have been lost at 4.1 Ga (Fig. 16). The surface water has experienced further loss to reach $\delta D = 5000 – 6000$ ‰ (Usui et al., 2012; Kurokawa et al., 2014; Mahaffy et al., 2014; Villanueva et al., 2015). An intermediate D/H ratio ($\delta D = 1000 – 2000$ ‰) recorded in impact glasses in Martian meteorites suggests that the majority of near-surface water might have been stored as subsurface ice isolated from the atmosphere (Usui et al., 2015; Kurokawa et al., 2016; Grimm et al., 2017).

On the other hand, if one assume for a Noachean ocean volume of about 500 m global equivalent layer (GEL) or 0.05 times the terrestrial ocean mass ($M_{TO}$) (Di Ahille and Hynek, 2010), the water loss before 4.1 Ga is > 160 m GEL, which might exceed the integrated oxygen escape (Jakosky et al., 2018). This imbalance between H and O escape drives the oxidation of the Martian surface environment (Lammer et al., 2003a; Zahnle et al., 2008).

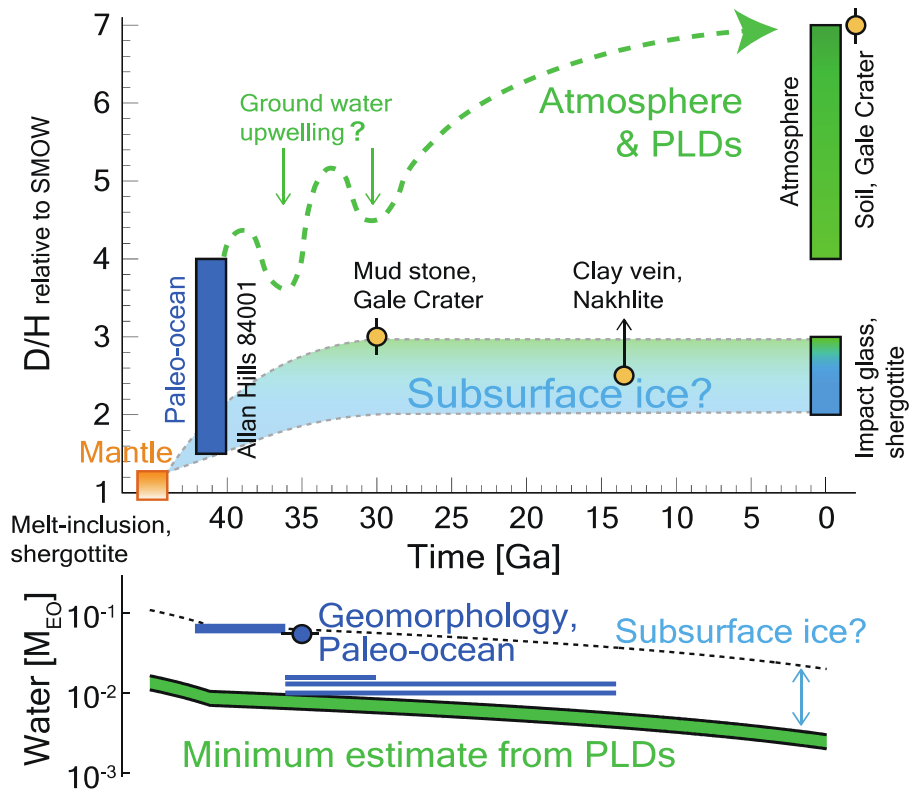

**Fig. 17.** Evolution of D/H (upper panel) and water volume (lower panel) on Mars (modified after Kurokawa et al. 2014, 2016; Usui et al., 2017). Data for D/H are from Usui et al. (2012, 2015, 2017), Boctor et al. (2003), Greenwood et al. (2008), Webster et al. (2013), Mahaffy et al. (2015), Villanueva et al. (2015), Krasnopolsky (2015), and Aoki et al. (2015). Water volume was computed by assuming Rayleigh fractionation with $f = 0$.

That Mars lost large amounts of water was also suggested by Lammer et al. (2003b), who estimated the present and past $H_2O$-ice reservoirs, which are and were in exchange with the atmosphere, by using the observed D/H ratio in the atmospheric water vapour, measured D/H ratios in Martian SNC meteorites and D/H isotope ratios based on a study regarding asteroid and cometary water delivery to early Mars by Lunine et al. (2003).

These authors estimated the present $H_2O$-ice reservoir to be equivalent to about 11 – 27 m GEL. From this obtained range, depending on atmospheric escape of D and H a reservoir exchangeable with the atmosphere on Mars 3.5 Ga equivalent to a global ocean with a thickness of between $\approx$ 17 to 61 m was found. If one uses these estimated water reservoirs 3.5 Ga ago, Lammer et al. (2003b) found water reservoirs affected by isotopic fractionation equivalent to a global ocean with a depth of between 97 - 115 m if 1.2 times the terrestrial sea water ratio was the initial isotope ratio in Mars' $H_2O$ 4.5 Ga as (Lunine et al., 2003). The difference with respect to the initial volume of about 0.05 $M_{TO}$ (Di Ahille and Hynek, 2010) may have escaped unfractionated to space between 3.5 - 4.5 Ga and/or partly stored in the subsurface (Orosei et al. 2018) (see also Fig. 17).

**8. 5 Redox state of the Martian atmosphere in relation to the atmospheric escape**

The preferential loss of hydrogen into space causes the oxidation of the Martian surface (e.g. Lammer et al., 2003a). Additionally, carbon escape may have dominated against oxygen escape on early Mars (Tian et al., 2009; Amerstorfer et al., 2017) which may have also oxidized Mars' surface. If this is the case, the redox state of the early Martian atmosphere could be more reducing than modern Mars. Thus, it is worth considering the possibility of an early reducing Martian atmosphere.

First, the outgassing composition of presumably early Martian volcanos could have been more reducing than Earth's since the oxygen fugacity of the Martian mantle could be up to 3 log-units lower than of the terrestrial mantle and possibly close to the IW-buffer especially during its early period (Wadhwa, 2001; Herd et al., 2002; Righter et al., 2008; McSween et al., 2009). When assuming a low oxygen fugacity (log $f_{O2}$ = IW+1), the outgassing composition should be roughly $H_2 : H_2O : CO : CO_2$ = 10 : 1 : 1 : 1 (Zolotov, 2003), which is much more reducing than Earth's present volcanic flux. Although the oxygen fugacity of the Martian crust may vary and could be higher than in the mantle (Wadhwa, 2001), the Martian basaltic volatiles could be dominated by $H_2$ and CO relative to $CO_2$ based on a thermodynamic calculation of Gaillard and Scaillet (2014). Given the $H_2$-rich volcanic flux, the early Martian atmosphere should have been very reducing and possibly rich in CO (Sholes et al., 2017). This is due to $CO_2$ continuously being photolyzed into CO, whereas the oxidation of CO back into $CO_2$ is not efficient if the volcanic flux provides sufficient reducing agents into the atmosphere. This might then lead to a CO-runaway (Yung and DeMore, 1998; Zahnle et al., 2008).

The reducing early Martian atmosphere may provide a solution to explain why the atmospheric $CO_2$ of Mars ($\delta^{13}C_{PDB}$ = +46 ± 4‰; Webster et al., 2013) is so enriched in $^{13}C$ compared to that of the Martian mantle ($\delta^{13}C_{PDB}$ = -25 ± 5‰; Wright et al., 1992). Galimov (2000) originally proposed that $CO_2$ could have been a minor component in the early Martian atmosphere as compared with $CH_4$ or CO from a thermodynamic point of view. $CH_4$ and CO are depleted in $^{13}C$ and preferentially escaped into space, whereas the $^{13}C$-enriched $CO_2$ is ultimately deposited as carbonate. Also, as mentioned above, Hu et al. (2015) argue that photodissociation of CO may significantly induce preferential loss of $^{13}C$.

In these scenarios, the atmospheric $^{13}C$-enrichment is caused by preferential escape of $^{13}C$ into space though time. Based on the isotopic studies of carbonate globules in ALH84001 (Eiler et al., 2002; Halvey et al., 2011), the enrichment in $^{13}C$ may have possibly occurred before 3.9 Ga, when the geomagnetic field may have still existed and according to Lillis et al., (2013) prohibited carbon escape into space via pick-up-ion sputtering induced by the solar wind. On the other hand, C atoms most likely escaped efficiently during this period by thermal escape (Tian et al., 2009; Amerstorfer et al., 2017). Moreover, due to the higher EUV flux of the young Sun early Mars upper atmosphere expanded to larger distances so that the intrinsic magnetic field may not have shielded exospheric particles against escape. A reducing atmosphere (a smaller $CO_2$ mixing ratio) leads to a higher temperature in the upper atmosphere due to inefficient cooling and therefore allows efficient escape (Kulikov et al., 2007; Tian et al., 2009; Johnstone et al., 2018). Further experimental and model studies are required to test the reducing early Martian atmosphere scenario.

An additional alternative mechanism for early $^{13}C$-enrichment recorded in ALH84001 than atmospheric escape is isotopic fractionation by solar UV photodissociation of $CO_2$, which may cause significantly large carbon isotopic fractionation of about 200 ‰ based on theoretical calculation of absorption spectra of the $CO_2$ isotopologues (Schmidt et al., 2013). Such a process is also relevant if Mars early atmosphere would have been lost during younger ages compared to that of ALH84001. Although the effect has not yet been tested by laboratory experiment, solar UV could dissociate $^{12}CO_2$ preferentially, and thus provide $^{13}C$-enriched carbonate precipitating from the remaining $CO_2$. In this case highly $^{13}C$-depleted CO is produced in the atmosphere. Under this reducing condition and the presence of water vapor,

CO could be converted mainly into formaldehyde and removed from the atmosphere (Bar-Nun and Hartman, 1978; Pinto et al., 1980; Bar-Nun and Chang, 1983). If this scenario is correct, then $^{12}$C preferentially deposited as organics into sediment, but did not escape into space.

## 9 Conclusion

We discussed the relevance of various loss processes that led most likely to today's atmospheric Ar, and Ne noble gas isotope anomalies and the depletion of moderately volatile rock-forming elements on the terrestrial planets. A complex interplay between the young Sun's EUV flux, the accreted protoplanetary mass within the disk lifetime, a possible accumulated primordial $H_2$-dominated atmosphere, the impacting bodies that were involved in the accretion before and after disk dispersal, as well as the depletion of volatile elements from planetary embryos and the final protoplanets led eventually to today's measured ratios. The discovery of many low mass exoplanets that have captured a primordial $H_2$-dominated atmosphere that never escaped to space during billions of years indicates that the accretion of smaller planets is also a fast process. One finds from early atmosphere evolution models that measured atmospheric $^{36}$Ar/$^{38}$Ar, $^{20}$Ne/$^{22}$Ne and $^{36}$Ar/$^{22}$Ne isotope ratios can be reproduced if proto-Venus and Earth were released from the solar nebula at ≈ 4 Myr with a small captured $H_2$-dominated envelope and protoplanetary masses of ≈ 0.85 – 1.0 $M_{Venus}$ and 0.53 – 0.58 $M_{Earth}$, respectively, and. From the isotopic analysis of evolution models one finds that post-$H_2$-envelope impactors with an admixture of ≥ 70% carbonaceous chondrites and ≤ 30% dry ureilite-like material reproduces Earth's present atmospheric $^{36}$Ar/$^{22}$Ne isotope ratios at best. These late stage impactors, however, have to be depleted in Ar, Ne and moderately volatile elements like K. As long as the accreting proto-planet were embedded in the nebula or surrounded by an the escaping nebula-based $H_2$-dominated primordial atmosphere the solar-like Ar and Ne isotope abundance dominated to any accreted, or from the underlying magma ocean outgassed composition/ratios. Because of this, one can only determine from reproduction attempts of today's atmospheric Ne and Ar the accreted material composition after the primordial atmosphere was lost. If EN-like material was involved in the accretion of proto-Earth as suggested by Dauphas (2017), then this material should have been delivered to early Earth's accretion as long as the growing proto-Earth was surrounded by nebula gas.

Impact erosion and outgassing of volatiles from magma pools and oceans and subsequent escape to space should have depleted the planetary embryos that acted in the final accretion of the early terrestrial planets. Latest reproduction attempts of atmospheric $^{36}$Ar/$^{38}$Ar and $^{20}$Ne/$^{22}$Ne isotope ratios and bulk K/U ratios on Venus agree with the findings related to early

Earth's evolution that the young Sun was less active than a moderate rotator und EUV fluxes corresponding to a faster rotating young Sun were very unlikely.

Thus, new precise measurements by future Venus' space missions of atmospheric Ar, Ne, Kr, and Xe noble gas isotope ratios and more data of moderately volatile elements distributed over a wide surface area would give a precise evolution scenario of Venus and would further help to constrain the Sun's history. For terrestrial exoplanets one can expect that different accretion scenarios and different radiation fields of their host stars will result in different heat producing elements and hence in different tectonic regimes (see also O'Neill et al., 2019; this issue), which can have tremendous implication for the planet's evolution and habitability.

From D/H analysis, one finds that Mars lost a huge fraction of its initial water inventory during the first 500 Myr after its origin. The low gravity and high EUV fluxes of the young Sun led also to high escape rates of its bulk atmosphere. After the EUV flux of the young Sun decreased at $\approx$ 4–4.1 Ga, volcanic outgassing and impacts could have resulted in the build-up of a denser atmosphere of $\leq$ 1 bar, which is also constrained by analysis of $^{36}Ar/^{38}Ar$ and $^{13}C/^{12}C$ data. Analysis of the atmosphere enriched in $^{13}C$ isotopes compared to Mars' mantle may be an indication that early Mars was more reduced as today. $CO_2$ could have been a minor component of the early Martian atmosphere as compared with CO or $CH_4$ from a thermodynamic point of view. Both molecules are depleted in $^{13}C$ and preferentially escaped into space, whereas $^{13}C$-enriched $CO_2$ is ultimately deposited as carbonate. $CO_2$ could have been released from surface reservoirs during periods of climate change or impacts and may have modified the atmospheric surface pressure several times during the planet's history. Since $\approx$ 3.5 Ga, when the young Sun's EUV flux decreased to lower values, Mars' atmospheric evolution until today was determined by a complex interplay of ion escape, atmospheric sputtering and suprathermal hot atom escape, carbonate precipitation, and serpentinization that led to today's low surface pressure.

**Acknowledgements** We acknowledge support by the Austrian Fonds zur Förderung der Wissenschaftlichen Forschung, Nationales Forschungs Netzwerk (FWF NFN) project S116-N16 and the subprojects S11603-N16, S11604-N16, S11606-N16, S11607-N16 and S11608-N16. H. Lammer, M. Leitzinger and P. Odert acknowledge support of the FWF projects P27256-N27 and P30949-N36. M. Benedikt and H. Lammer acknowledge support from the Austrian Forschungsförderungsgesellschaft (FFG) project RASEN. L. Fossati acknowledge also the FFG project "TAPAS4CHEOPS" P853993. H. Kurokawa acknowledges support from JSPS KAKENHI Grant 17H01175, 17H06457, 18K13602, 19H01960, and 19H05072.